\documentclass[
	fontsize=10pt,          
	numbers=noenddot,    	
    parskip=half,        	
    listof=totoc,        	
    bibliography=totoc,  	
	headsepline=true,       
	footsepline=false, 		
    DIV=12                	
]{scrartcl}

\usepackage[utf8]{inputenc}
\usepackage[T1]{fontenc}
\usepackage[english]{babel}

\usepackage{graphicx}

\usepackage{amsmath, amsfonts, amssymb, bbm, bm, mathtools}

\usepackage{apacite} 
\usepackage{natbib}

\usepackage[hidelinks, breaklinks]{hyperref} 
\usepackage{url}
\usepackage[doublespacing]{setspace} 

\usepackage{pifont}   

\usepackage{wrapfig}

\usepackage{nicefrac}

\usepackage{pdfpages}

\usepackage{rotating}

\usepackage[ruled,vlined]{algorithm2e}  

\usepackage{scrlayer-scrpage}
\clearscrheadings                   
\automark{section}                  
\ihead{Comparing models for ordinal outcomes}				
\ohead{\pagemark}					
\ifoot{}							
\ofoot{}							

\usepackage{booktabs} 	

\usepackage{csquotes}

\usepackage{booktabs}
\usepackage{caption}
\usepackage[export]{adjustbox} 
\usepackage[left=1truein,right=1truein,top=1truein,bottom=1truein]{geometry}
\usepackage{abstract}

\usepackage[graphicx]{realboxes}
\usepackage{pdflscape}

\usepackage{multirow}

\usepackage{makecell}

\begin{document}

\thispagestyle{empty}
\newgeometry{top=2.5in,bottom=1in,left=1in,right=1in}

\vspace{15em}

\begin{center}
\Large{\textbf{A Simulation Study to Compare Inferential Properties when Modelling Ordinal Outcomes: The Case for the (Plain but Robust) Proportional Odds Model}}    
\end{center}

\vspace{5em}

\begin{center} 
    Stefan Inerle$^{1}$, Markus Pauly$^{1,2}$, Moritz Berger$^{3}$
    
    \vspace{2em}
    
    \footnotesize{
        $^{1}$Department of Statistics, TU Dortmund 
        University\\
        $^{2}$Research Center Trustworthy Data Science and Security, University Alliance Ruhr (UA Ruhr)}\\
        $^{3}$Core Facility Biostatistics, Central Institute of Mental Health, Medical Faculty Mannheim, Heidelberg University, Mannheim, Germany       
\end{center}

\vspace{4em}
\begin{center} 
    \textbf{Autor Note}

Stefan Inerle \includegraphics[height=10pt]{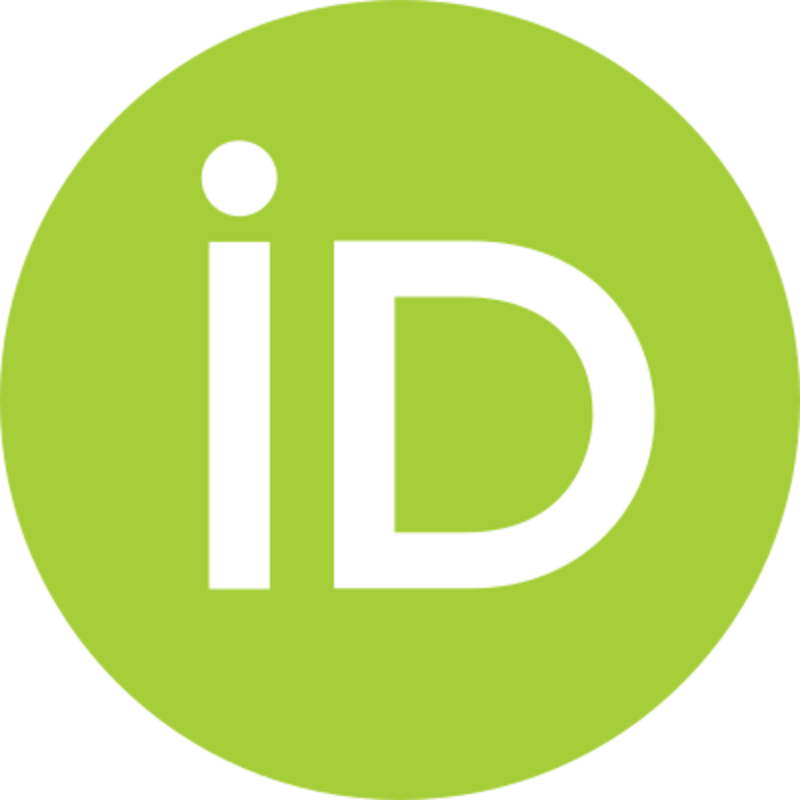} \url{https://orcid.org/0009-0002-7259-4760} \\
Markus Pauly \includegraphics[height=10pt]{orcid_logo.png} \url{https://orcid.org/0000-0002-0976-7190} \\
Moritz Berger \includegraphics[height=10pt]{orcid_logo.png} \url{https://orcid.org/0000-0002-0656-5286} 
\vspace{3em}

Correspondence concerning this article should be addressed to Stefan Inerle, Department of Statistics, TU Dortmund University, Vogelpothsweg 87, 44221 Dortmund Germany. Email: stefan.inerle@tu-dortmund.de
\end{center}

\newpage

\restoregeometry 

\begin{abstract}
Ordinal measurements are common outcomes in studies within psychology, as well as in the social and behavioral sciences. Choosing an appropriate regression model for analysing such data poses a difficult task. 
This paper aims to facilitate modeling decisions for quantitative researchers by presenting the results of an extensive simulation study on the inferential properties of common ordinal regression models: the proportional odds model, the category-specific odds model, the location-shift model, the location-scale model, and the linear model, which incorrectly treats ordinal outcomes as metric. 
The simulations were conducted under different data generating processes based on each of the ordinal models and varying parameter configurations within each model class. We examined the bias of parameter estimates as well as type I error rates ($\alpha$-errors) and the power of statistical parameter testing procedures corresponding to the respective models. 
Our findings reveal several highlights. For parameter estimates, we observed that cumulative ordinal regression models exhibited large biases in cases of large parameter values and high skewness of the outcome distribution in the true data generation process. Regarding statistical hypothesis testing, the proportional odds model and the linear model showed the most reliable results. Due to its better fit and interpretability for ordinal outcomes, we recommend the use of the proportional odds model unless there are relevant contraindications. 
\\

\hspace{10mm} \textit{Keywords}: Ordinal Regression, Regression Models, Odds, Estimation Bias,  $\alpha$-Error, Power
\end{abstract}
\newpage



\section{Introduction}

Handling ordinal data is a common challenge in various scientific fields, especially in psychology \citep[e.g.][]{joshiLikertScaleExplored2015}, but also in other disciplines such as medicine \citep[e.g.][]{sullivanAnalyzingInterpretingData2013}. This is particularly true for regression problems, which we study in this paper. A key challenge is the selection of an appropriate model and method for analyzing ordinal outcomes. 
To this end, we first need to decide whether we want to treat the outcome as metric or not. 

Many authors recommend the use of ordinal models to emphasize that treating ordinal outcomes as metric can lead to interpretational problems \citep[e.g.][]{brunner2018rank} or inferential errors \citep[e.g.][]{liddellAnalyzingOrdinalData2018}, 
for example when means are close to the end of the ordinal scale. 
If we choose to use an ordinal model, the next step is to decide which one. \cite{tutzOrdinalRegressionReview2022} provides a comprehensive review of ordinal regression models, while \cite{burknerOrdinalRegressionModels2019} offer a theoretically motivated recommendation to select an ordinal model based on how well it reflects the theoretically assumed relations in the data. 

In contrast, \cite{kniefViolatingNormalityAssumption2021} argue that Gaussian models are remarkably robust, even for categorical data. They additionally stress that more sophisticated methods may carry inherent risks that quantitative researchers might not fully understand or be aware of. Moreover, many applied researchers favor simpler Gaussian linear models. The latter is also reflected in the literature review presented in \cite{liddellAnalyzingOrdinalData2018}, in which  
all reviewed studies involving ordinal outcomes used metric models to analyze them.

To underpin this discussion more closely, and in preparation for our paper, we conducted a literature review using PubPsych. The goal was to gain a clearer understanding of which ordinal models are used for ordinal outcomes in empirical research. To this end, we searched PubPsych (\url{https://pubpsych.zpid.de/pubpsych/}) using the search terms ``regression'' and either ``ordinal'' or ``ordered'', and only included studies published between 2014 and 2023 for which we had access to the full text. \newline
After excluding all papers that did not explicitly report the use of an ordinal model, we were left with 458 papers. Among these, 6 papers used two different methods, and one paper used 4 different methods. In 11 papers, it was unclear how or where an ordinal model was used. 50 papers implemented some form of mixed model, and 4 used Bayesian methods, which both extend beyond the scope of this study. 
Among the remaining 396 papers, the most frequent model was the proportional odds model \citep{mccullaghRegressionModelsOrdinal1980, tutz2011regression} appearing in 205 (52\%) papers. 
Additional identified models were the partial proportional odds model \citep{petersonPartialProportionalOdds1990} in 14 papers (3.5\%) and the category-specific odds model \citep{petersonPartialProportionalOdds1990, tutz2011regression} in 2 papers (0.5\%). Other papers used different link functions than the logit link: We observed 12 papers with a probit link and 5 with a negative log-log link. 23 papers did not state the link function explicitly. 
136 (34.3\%) papers used the logit link function, but did not explicitly state which model was used. Out of these papers, 105 papers reported one odds ratio per covariate, 14 reported one regression weight per covariate, and 7 reported both, while 10 reported none. \newline
In summary, our literature review reveals that many empirical studies do not specify the applied ordinal model. Among those that do, the proportional odds model was the most frequently used ordinal regression approach.

While such theoretical and empirical discussions are valuable, simulation studies can provide deeper insight and guide practitioners to select the most suitable models for the analysis of ordinal outcomes. Although there are several simulation studies that compare the predictive performance of various ordinal regression models
\cite[e.g.][]{janitzaRandomForestOrdinal2016,hornungOrdinalForests2020,tutzOrdinalTreesRandom2022,buczakOldGoldNew2024,buczak2025frequency}, simulation studies regarding inferential analyses are scarce. 
Closing this gap to provide concrete recommendations for inference in parametric ordinal regression models for ordinal outcomes is the primary goal of our paper. To maintain a clear focus and ensure comparability, we excluded models that incorporate regularization \cite[e.g., via mixed modeling or Bayesian approaches, for details see][]{friedrich2023regularization} from our study. 
The chosen models for comparison, which are formally introduced in the next section, include the proportional odds model \citep{mccullaghRegressionModelsOrdinal1980, tutz2011regression}, the category-specific odds model \citep{petersonPartialProportionalOdds1990, tutz2011regression}, the location-shift model \citep{tutzSeparatingLocationDispersion2017}, and the location-scale model \citep{mccullaghRegressionModelsOrdinal1980}. The location-shift model is a relatively recent development \citep{tutzSeparatingLocationDispersion2017}
which may be the reason why it did not appear in our literature review.
It was invented as an alternative to the location-scale model with a software implementation in R \citep{bergerOrdDispSeparatingLocation2020}.
For all these models, we investigate parameter estimations (bias) and tests ($\alpha$-error and power). 
To contribute to the discourse on whether choosing metric models may lead to worse results, we also include the linear model for the investigation of parameter tests.
\section{Methods}

Our simulation study compares the bias of parameter estimates as well as $\alpha$-error and power of statistical tests. In this section, we sketch the different models considered and describe our experimental design. We denote $Y \in \{1, \hdots, k\}$ as the ordinal outcome variable with $k$ categories and $\mathbf{X}=(X_{1},\hdots,X_{p})^\top$ as the vector of covariates.

\subsection{Regression Models}

Treating $Y$ as metric, the \textit{Gaussian model} (LM) with linear predictor function \citep[e.g.][]{hastieElementsStatisticalLearning2009b} is given by 
\begin{align}
\label{LinReg}
Y = \beta_{0} + \sum_{j = 1}^p X_{j} \beta_j +\varepsilon= \beta_{0} + \mathbf{X}^\top\boldsymbol{\beta}+\varepsilon\,,\quad \varepsilon \sim N(0,\sigma^2)\,,
\end{align}
where $\beta_0 \in \mathbb{R}$ is the intercept coefficient, $\boldsymbol{\beta} = (\beta_1, \hdots, \beta_p) \in \mathbb{R}^p$ is the vector of real-valued regression coefficients and $\sigma^2 \in \mathbb{R}_{>0}$ denotes the error variance. 

Treating $Y$ as ordinal, a widely used tool is the class of \textit{cumulative regression models} \citep{mccullaghRegressionModelsOrdinal1980, tutz2011regression}. It is derived from an underlying latent continuous variable $Y^*$ which follows a regression model $Y^*=-\mathbf{X}^\top\boldsymbol{\beta}+\varepsilon$, where $\varepsilon$ is a noise variable with continuous distribution function $F(\cdot)$. Instead of $Y^*$, we only observe the coarser categorical version $Y$ determined by $(Y=r) \Leftrightarrow(\theta_{r-1} <Y^* \leq \theta_{r})\,,$ where $-\infty=\theta_0 < \theta_1 < \hdots < \theta_k = \infty$ are thresholds on the latent scale. We then obtain the cumulative model 
\begin{align*}
\mathrm{P}(Y\leq r \mid \mathbf{X}) = \mathrm{F}(\theta_r+ \mathbf{X}^\top \boldsymbol{\beta})\,,\quad r=1,\hdots,k-1\,.
\end{align*}
Since $P(Y\leq k | \mathbf{X})=1$, we don't explicitly model it. The most popular choice for $F(\cdot)$ according to our literature review is the logistic distribution function yielding the \textit{proportional odds model} (PO)
\begin{align}
\label{Prop_odds_expit}
\mathrm{P}(Y\leq r \mid \mathbf{x}) = \frac{\exp{(\theta_r + \mathbf{X}^\top \boldsymbol{\beta})}}{1 + \exp{(\theta_r + \mathbf{X}^\top \boldsymbol{\beta})}}\,,\quad r=1,\hdots,k-1\,.
\end{align}
It uses the same global linear predictor function for all categories $r$ and allows for a simple interpretation of the regression coefficients: Let $\gamma(r|\mathbf{X})=\mathrm{P}(Y\leq r \mid \mathbf{X})/\mathrm{P}(Y > r \mid \mathbf{X})$ denote the cumulative odds. Then the proportion of cumulative odds for two sets of covariates $\mathbf{X}$ and $\tilde{\mathbf{X}}$ is given by 
\begin{align*}
\frac{\gamma(r|\mathbf{X})}{\gamma(r|\tilde{\mathbf{X}})}=\exp\left((\mathbf{X}-\tilde{\mathbf{X}})^\top\boldsymbol{\beta}\right)\,,
\end{align*}
which does not depend on category $r$. Thus, if $X_j$ increases by one unit, the cumulative odds change by the factor $\exp(\beta_j)$. 

The proportional odds model can be extended to contain category-specific effects \citep{petersonPartialProportionalOdds1990, tutz2011regression} by assuming that each category has a distinct vector of regression coefficients. This leads to  the \textit{category-specific odds model} (CSO) 
\begin{align}
\label{Cat_odds_expit}
\mathrm{P}(Y\leq r \mid \mathbf{X}) = \frac{\exp{(\theta_r + \mathbf{X}^\top \boldsymbol{\beta}_r)}}{1 + \exp{(\theta_r + \mathbf{X}^\top \boldsymbol{\beta}_r)}}\,,\quad r=1,\hdots,k-1\,,
\end{align}
which contains parameter vectors $\boldsymbol{\beta}_r^\top=(\beta_{1r},\hdots,\beta_{pr})$ that might differ across categories. The model in~\eqref{Cat_odds_expit} is also referred to as \textit{non-proportional odds model}. It provides high flexibility, however, postulates that $\theta_r + \mathbf{X}^\top \boldsymbol{\beta}_1 \leq \hdots \leq \theta_{k-1} + \mathbf{X}^\top \boldsymbol{\beta}_{k-1}$ for all values of $\mathbf{X}$. This strongly restricts the possible values of covariates, which was also one of the major issues that caused problems in our simulation.  

If not all parameters vary over categories in \eqref{Cat_odds_expit}, we observe two types of covariates: those with a global effect, for which $\beta_{j1}=\hdots=\beta_{j,k-1}=\beta_j$, and those with category-specific effects. \citet{petersonPartialProportionalOdds1990} refer to this model as the partial proportional odds model. To distinguish between covariates with global and category-specific effects, a selection strategy is required. In our literature review, we found that some researchers apply a selection algorithm implemented in STATA, where category-specific effects are successively collapsed using a backward selection procedure based on p-values of Wald tests.  

Another alternative for the proportional odds model, which allows for differing variability in subgroups of the population -- so-called dispersion effects -- is the \textit{location-shift model} (LSH) introduced by \cite{tutzSeparatingLocationDispersion2017}. Let $\boldsymbol{Z}=(Z_1,\hdots,Z_q)^\top$ be an additional vector of covariates for which dispersion effects are assumed. Then the location-shift model has the form
\begin{align}
\label{loc_shift_expit}
\mathrm{P}(Y\leq r \mid \mathbf{X}) = \frac{\exp{(\theta_r + \mathbf{X}^\top \boldsymbol{\beta} +(r-k/2)\,\mathbf{Z}^\top \boldsymbol{\gamma})}}{1 + \exp{(\theta_r + \mathbf{X}^\top \boldsymbol{\beta} +(r-k/2)\,\mathbf{Z}^\top \boldsymbol{\gamma})}}\,,\quad r=1,\hdots,k-1\,,
\end{align}
where $\boldsymbol{\gamma}=(\gamma_1,\hdots,\gamma_q) \in \mathbb{R}^q$ is the vector of dispersion parameters. The model in \eqref{loc_shift_expit} additionally contains the scaled shifting term $(r-k/2)\mathbf{Z}^\top \boldsymbol{\gamma}$, which determines the shifting of thresholds and reflects the tendency to low or high categories. The scaling factor $(r-k/2)$ is an additional weight, which affects that the shifting of thresholds is proportional to the distance from the middle threshold, and hence the intervals between all thresholds are widened by the same value. The location-shift model can be seen as a category-specific odds model with specific constrains: Let $\boldsymbol{X}=\boldsymbol{Z}$, then one has $\theta_r + \mathbf{X}^\top \boldsymbol{\beta} +(r-k/2)\,\mathbf{X}^\top \boldsymbol{\gamma}=\theta_r + \mathbf{X}^\top\left(\boldsymbol{\beta} +(r-k/2)\,\boldsymbol{\gamma}\right)=\theta_r + \mathbf{X}^\top\boldsymbol{\beta}_r$. Because the proportional odds model is a submodel of the location-shift model, the following nested structure holds
\[
\textrm{proportional odds model} \subset \textrm{location-shift model} \subset \textrm{category-specific  odds model}\,,
\]
which is outlined in detail in \citet{tutzSparserOrdinalRegression2022}.

An alternative to account for additional dispersion effects is obtained by assuming  $Y^*=-\mathbf{X}^\top\boldsymbol{\beta}+\tau_{\boldsymbol{Z}}\varepsilon$
for the latent regression model, where $\tau_{\boldsymbol{Z}}$ is the variance parameter depending on $\boldsymbol{Z}$. Then the cumulative model is called \textit{location-scale model} (LSC) and has the form \citep{mccullaghRegressionModelsOrdinal1980}
\begin{align}
\label{loc_scale_expit}
\mathrm{P}(Y\leq r \mid \mathbf{X}) = \exp{\Bigl(\frac{\theta_r + \mathbf{X}^\top \boldsymbol{\beta}}{\tau_{\boldsymbol{Z}}}\Bigl)} \cdot \biggl(1 + \exp{\Bigl(\frac{\theta_r + \mathbf{X}^\top \boldsymbol{\beta}}{\tau_{\boldsymbol{Z}}}\Bigl)} \biggl)^{-1}\,,\quad r=1,\hdots,k-1\,,
\end{align}
where $\tau_{\boldsymbol{Z}}=\exp(\boldsymbol{Z}^\top\boldsymbol{\gamma})$. This model is non-linear and therefore cannot be embedded within the framework of multivariate generalized linear models. If $\boldsymbol{X}$ and $\boldsymbol{Z}$ are distinct, the interpretation of the $\boldsymbol{X}$ variables in terms of cumulative odds for the location-shift model as well as for the location-scale model is the same as in the proportional odds model described above. 

In our simulation study, we include the following five models: the linear model (LM), the proportional odds model (PO), the category-specific odds model (CSO), the location-shift model (LSH), and the location-scale model (LSC). However, the selection of covariates with global and category-specific effects, as in the partial proportional odds model, is out of the scope of this paper.

\subsection{Simulation}
In this study, we generated data from the proportional odds model, the category-specific odds model, the location-shift model, and the location-scale model. 
The number of observations for this simulation study was set to $n \in \{250, 500, 1000\}$, as we recognized in a pilot study that lower sample sizes will not lead to relatively stable models and higher values may be unrealistic in many applications. Furthermore, picking 250 observations as the lowest number made sure that in each scenario there were more observations than parameters to be estimated. We included either 5 or 35 covariates, where none, one, or four of them were truly informative (with a coefficient different from zero). The number of categories $k$ of the ordinal outcome was chosen as 3, 5, or 7. All covariates were independently normally distributed with mean zero and standard deviation 0.5. 
For models with constant location parameters $\beta$, we used parameter values $0$ (non-informative) and $0.1, 0.2, 0.5, 1, 2$ (informative) for all categories. These values were also used for the dispersion parameters $\gamma$. All $\beta$ values were combined with $\gamma$ values of 0 and 1, and all $\gamma$ values were combined with $\beta$ values of 0 and 1. This procedure was chosen to include a situation where neither location nor dispersion had a true effect, two situations where one of them had a true effect once, and one situation where both of them had true effects larger than zero. For the category-specific odds model, we also set the values $\beta_r$ to $0, 0.1, 0.2, 0.5, 1$ and~$2$, but with varying signs so that there was not simply a constant effect. For five categories, we chose $\boldsymbol{\beta} = (-u, u, u, -u)$ and for seven categories $\boldsymbol{\beta} = (-u, 0, u, u, 0, -u)$. In the case of three covariates, the category-specific model can be transformed into a location-shift model. Therefore, no data was generated with the category-specific model for $k = 3$. In the case of four informative covariates, all of them were drawn with the same regression weights.\newline
In the cases where none of the covariates is informative, the thresholds $\boldsymbol{\theta}$ define the distribution of the outcome variable. Therefore, three different $\boldsymbol{\theta}$ vectors were implemented. 

\begin{figure}[h]
    \centering
    \includegraphics[width=0.5\textwidth]{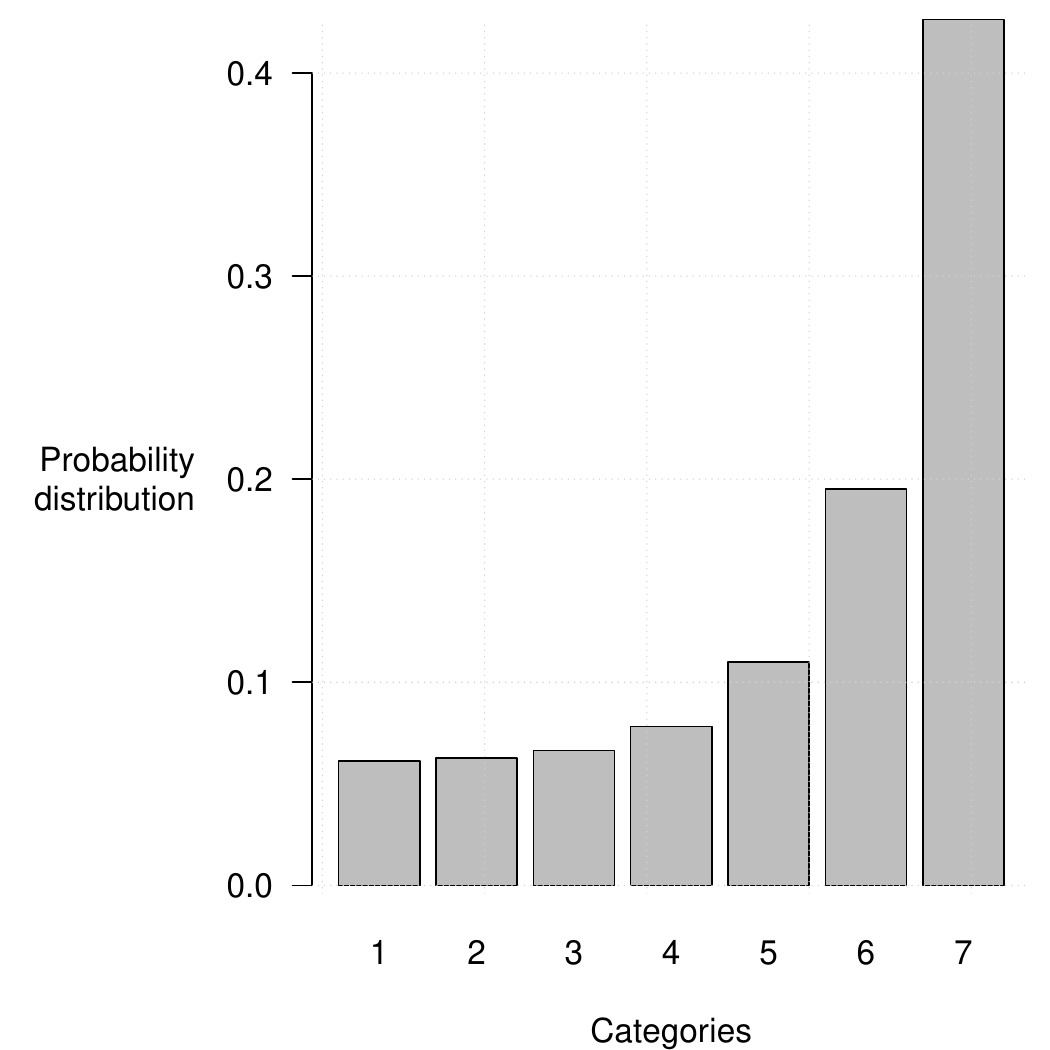}
    \caption{Probability distribution of the outcome variable for seven categories and no effect of the covariates when setting $\boldsymbol{\theta} = (-2.74, -1.96, -1.45, -1.00, -0.50, 0.29)^T$ (skewed setting).}
    \label{dist_2}
\end{figure}

In the first setting (uniform), the probabilities of the outcome variable were uniformly distributed
without the effects of the covariates. For the second setting (skewed), the true distribution was chosen so that at least six percent of the observations fell in each category. Additionally, higher values were exponentially more likely than lower values. For seven categories, this led to the values $\boldsymbol{\theta} = (-2.74, -1.96, -1.45, -1.00, -0.50, 0.29)^T$ and the distribution shown in Figure \ref{dist_2}. In the third setting (unstructured) for $\boldsymbol{\theta}$, we chose a distribution where the middle category and the lowest category were always a little more than twice as likely as all other categories, which were equally likely. When introducing effects for $\beta$ or $\gamma$ different from zero, the conditional outcome distributions change; nonetheless, the values for $\boldsymbol{\theta}$ were kept the same to achieve higher comparability between the simulation settings. Taking all combinations of varying simulation parameters, we obtained 4032 different simulation settings.

We performed 2000 replications for each setting and fitted the proportional odds model, the category-specific odds model, the location-shift model, and the location-scale model. For these models, the biases of the parameter estimates were calculated. Including the linear model as well, the proportion of significant parameter tests for each single parameter was calculated to achieve an estimation of the $\alpha$-error (considering the zero effects) and the power of the corresponding Wald-type tests with $\alpha = 0.05$ (considering the nonzero effects). 

\paragraph{Problems in the simulation}
When drawing data from the ordinal regression models, several problems appeared. In case a data set was drawn, where one outcome category included fewer than five observations, the data set was discarded and redrawn. This happened particularly often for data generation mechanisms with large dispersion, a low number of observations, a high number of outcome categories, and in the skewed distribution setting. \newline
As outlined in the previous section, for cumulative ordinal regression models, the condition $P(Y \leq r | \mathbf{X}) \leq P(Y \leq r+1 | \mathbf{X})$ must hold for all $r$. If observations were generated that did not fulfill this criterion, the observations were discarded and redrawn. This condition was often not met when drawing data from the category-specific model with high $\beta$ values and four informative covariates. \newline
For the category-specific model, there were 382 simulation settings, in which at least 20\% of the models did not converge and 639 simulation settings in which at least 5\% of the models did not converge. For the location-scale model, the respective numbers were 807 and 1150 simulation settings. Non-convergence in the category-specific model happened relatively often when there was a high number of covariates, a low number of observations, and large dispersion effects.
The location-scale model exhibited convergence difficulties when there was a high number of covariates, a low number of observations, few response categories, and large dispersion effects.  When the data was drawn from a location-scale model, there was large dispersion and low $\beta$ values, the location-shift model occasionally showed non-convergence in three settings where at least 20\% of the models did not converge and 105 settings where at least 5\% of the models did not converge.
More details concerning these convergence problems can be found in Appendix \ref{problems_appendix}.

In the following, results are only reported for those settings, where at least 100 out of the 2000 (i.e. $5\%$) simulation replications converged.

\subsection{R packages}
All calculations were performed in R, version 4.4.0 \citep{rcoreteamLanguageEnvironmentStatistical2015}. The package \texttt{VGAM} version 1.1-11 \citep{yeeVGAMPackageCategorical2010} was used for the proportional odds models and category-specific odds models. \texttt{ordDisp\texttt} version 2.1.1 \citep{bergerOrdDispSeparatingLocation2020} was used for the location-shift models and \texttt{ordinal} version 2023.12-4 \citep{christensenOrdinalRegressionModelsOrdinal2023} for location-scale models.\newline
Transparency in plots is achieved with the \texttt{scales} package version 1.3.0 \citep{wickhamScalesScaleFunctions2023}.

\newpage

\section{Results}
In this section, first, the bias of parameter estimates is presented and then the resulting $\alpha$-errors and power of the corresponding Wald tests follow. 

\subsection{Evaluation of the bias}
In this section, the average bias across all converged simulation replications is reported for the proportional odds model, the category-specific odds model, the location-shift model, and the location-scale model. Here, we report the fitted model that 
corresponds to the true data generating model to ensure that the true values of the regression parameters are known.

\begin{figure}[h]
    \centering
    \includegraphics[width=1\textwidth]{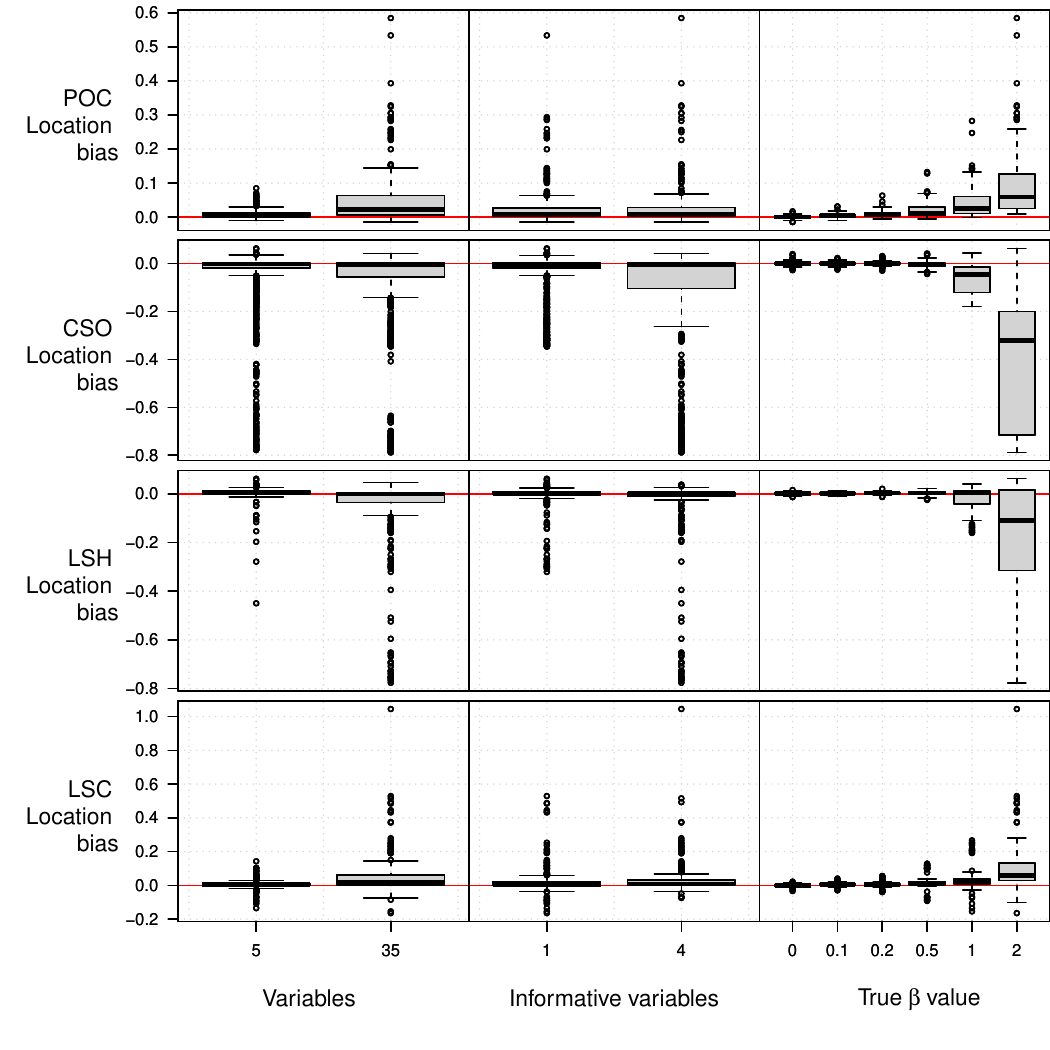}
    \caption{Influence of simulation settings on the biases when drawing with the proportional odds model aggregated across all $n$, $k$, and all distribution settings. Except in their specific columns, boxplots were aggregated across all numbers of covariates, number of informative variables, and true location parameter values.}
    \label{bias_po}
\end{figure}

\subparagraph{Location parameters $\beta$}
When generating data from the proportional odds model, the location parameters are the same and therefore comparable for all the cumulative models of interest. Therefore, Figure~\ref{bias_po} displays the bias of the location parameter of the four models with the proportional odds model as the data generation process. The results display the most important influences on the bias aggregated across all other simulation parameters.

Figure~\ref{bias_po} shows that an increasing number of covariates lead to a larger bias for all models and a larger dispersion of the bias. In particular,  a higher number of informative covariates increased negative bias in the category-specific odds models and, to a lesser extent, in the location-shift model. In contrast, they increased positive bias in the proportional odds and the location-scale models. More informative covariates lead to a large dispersion in bias for the category-specific model. The biggest influence on location bias was the true value of the location parameters, with higher values resulting in larger absolute bias and larger dispersion of the bias, especially for the category-specific odds model and the location-shift model. 

Overall, we observed that the proportional odds model and the location-scale model skewed slightly towards larger estimates than the true value. On the other hand, for the category-specific model and the location-shift model, we observed slightly lower estimates than the true value. We observed that for all models, the distribution $\theta$ did not have a relevant influence on the biases. The number of categories was irrelevant for all models except for the category-specific model, where more categories increased the negative bias very slightly. As expected, all biases decreased with an increasing number of observations. However, even with 1000 observations, some bias remained.

When generating data with the location-shift model, we saw the same patterns for the location parameter as before, but the negative bias was slightly larger for large true location parameter values and there was more dispersion in the estimations (displayed in Appendix~\ref{further_biases}). Similarly, when generating data with the location-scale model, the same patterns as described above appeared, though with slightly larger dispersion in the estimations (displayed in Appendix~\ref{further_biases}). Larger dispersion parameter values also increased the bias and the dispersion of the bias of the location parameters of the location-shift and the location-scale model. 

In the same way, when data was generated using the category-specific odds model and the corresponding location parameter estimates were calculated, similar patterns as described above were observed, with slightly stronger dispersion in the estimations (displayed in Appendix~\ref{further_biases}).

\subparagraph{Dispersion parameters $\gamma$}
In this paragraph, the location-shift and location-scale models are each evaluated under their respective data-generating processes. This procedure ensures that we know the true value of the parameters and can compare the two methods concerning the bias of the dispersion parameters.

\begin{figure}[h!]
    \centering
    \includegraphics[width=\textwidth]{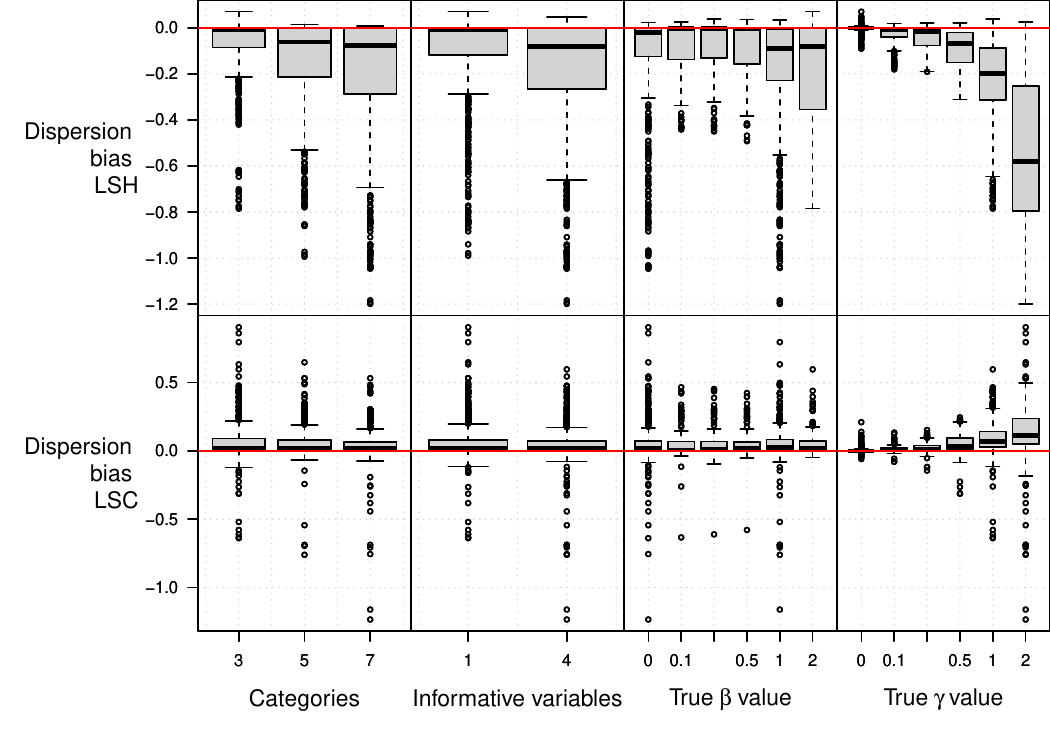}
    \caption{Influence of simulation settings on the dispersion biases when drawing with the respective model, including dispersion aggregated across all $n$, number of covariates, and all distribution settings. 
    Except in their specific columns, boxplots were aggregated across all numbers of categories $k$, number of informative variables, and true location parameter values, and true dispersion parameter values.}
    \label{bias_disp}
\end{figure}

For our simulation settings, we saw that the estimation of the dispersion parameter of the location-shift model underestimated the true value, while the location-scale model slightly overestimated the true value. We observed that for both models, the distribution parameter $\theta$ did not have a relevant influence on the biases. Looking at the number of covariates, for the location-shift model, the negative bias slightly increased, and for the location-scale model, the positive bias slightly increased. For both models, the biases slightly decreased with increasing number of observations. Nonetheless, with 1000 observations, there was still bias left. The influence of all other simulation parameters can be seen in Figure~\ref{bias_disp}. We observed that a higher number of categories increased the absolute value of the negative bias of the location-shift model, while it did not meaningfully change the bias in the location-scale model. The same effect could be seen for a higher number of informative covariates. Furthermore, we saw that with increasing true value of the location parameter, the absolute value of the bias of the dispersion parameter increased in the location-shift model, but not in the location-scale model. Lastly, we saw an increase in the absolute value of the bias with an increase in the true value of $\gamma$ in both models,
which was more pronounced in the location-shift model. 

\newpage

\clearpage

\newpage

\begin{figure}[h]
    \centering
    \includegraphics[width=1\textwidth]{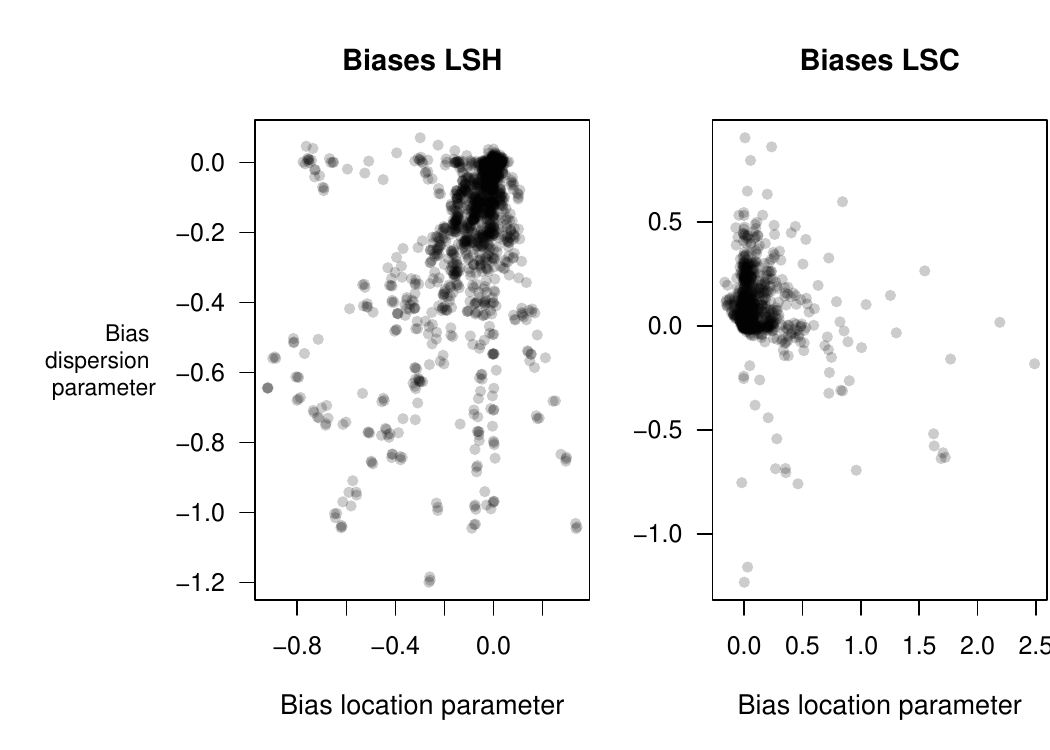}
    \caption{Relationship between the biases of the location parameter and the dispersion parameter in the location-shift and the location-scale model.}
    \label{relation_bias_disp}
\end{figure}

In Figure~\ref{relation_bias_disp}, we can see the relationship between the biases of the location parameter and the dispersion parameter for both the location-shift model and the location-scale model. The correlation between these biases for the location-shift model was 0.45 and -0.17 for the location-scale model, showing that the biases of the location-shift model were positively related.

\newpage 

\subsection{$\alpha$-Error}
The evaluation of the $\alpha$-error is done for hypothesis tests for the relevant parameter to be zero, where the true regression parameters are zero as well.

\paragraph{$\alpha$-Error when drawing from the proportional odds model}

\begin{figure}[h!]
    \centering
    \includegraphics[width=\textwidth]{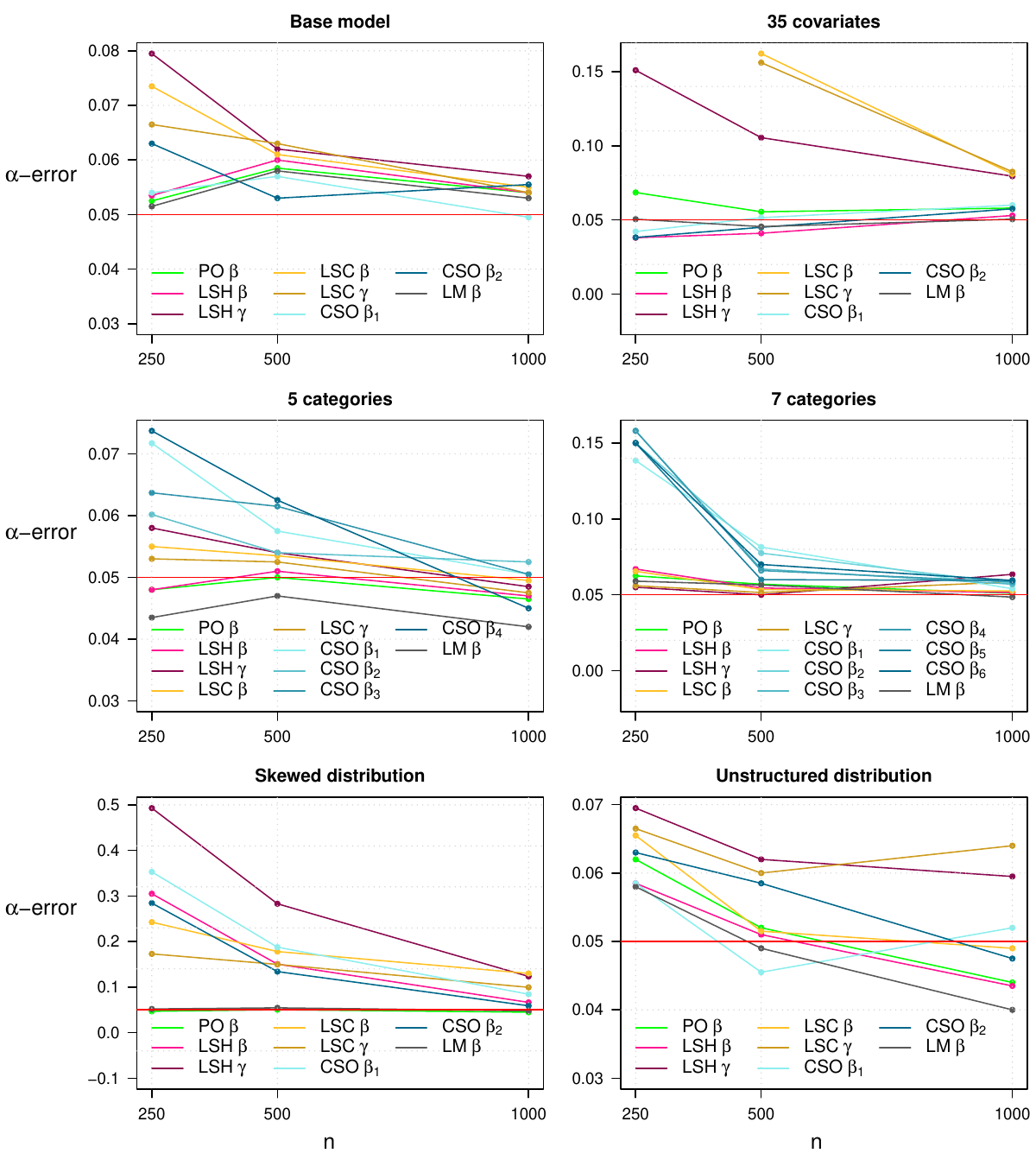}
    \caption{$\alpha$-error of the models when drawing with the PO model. The chosen base model has $p = 5$ covariates, one informative covariate, $k = 3$ categories, and the uniform distribution setting.}
    \label{alpha_po}
\end{figure}

In the following, we chose the simulation setting with $p = 5$ covariates, $k = 3$ categories, the uniform distribution, and data drawn from the proportional odds model with $\beta = 0$ (i.e., no informative covariate) and $\gamma = 0$ as the baseline model. We then displayed the $\alpha$-error for this model as well as for variations from this setting. We see in Figure~\ref{alpha_po} that in the baseline setting, the parameters of every model got close to adhering to the $\alpha$-level with an increasing number of observations. The dispersion parameter of the location-shift model showed the highest $\alpha$-error overall, with a value of about 0.08 for 250 observations. Additionally, both parameters of the location-scale model displayed relatively high $\alpha$-errors compared to the parameters of the other models. 

Figure~\ref{alpha_po} also shows how changing one simulation parameter in the data generation process changed the $\alpha$-error. When increasing the number of covariates, the tests for the linear model, the category-specific odds model, and the location parameter of the location-shift model got slightly more conservative, while the parameter tests of the proportional odds model got slightly more liberal. The dispersion parameter of the location-shift model showed considerably higher values, with about 0.15 at 250 observations. The location-scale model was most affected by the number of covariates, where tests for both parameters became too liberal, resulting in $\alpha$-errors of approximately 0.16 for 500 observations. For 250 observations, only 12 out of 2000 location-scale models converged. \newline
Increasing the number of categories had minimal impact on the relative performance of the models compared to the baseline setting. However, the category-specific odds model became increasingly liberal, especially for a lower number of observations. For seven categories, six parameters needed to be estimated, and all of them showed an $\alpha$-error of about 0.15 for 250 observations.\newline
Changing the distribution from uniform to skewed lead to a large increase in the $\alpha$-errors of the location-shift, the location-scale, and the category-specific odds model, especially for a lower number of observations. For 250 observations, we saw an $\alpha$-error of about 0.5 for the tests of the dispersion parameter of the location-shift model. The unstructured distribution performed similarly to the baseline model. 

\newpage

\clearpage

\begin{figure}[h]
    \centering
    \includegraphics[width=1\textwidth]{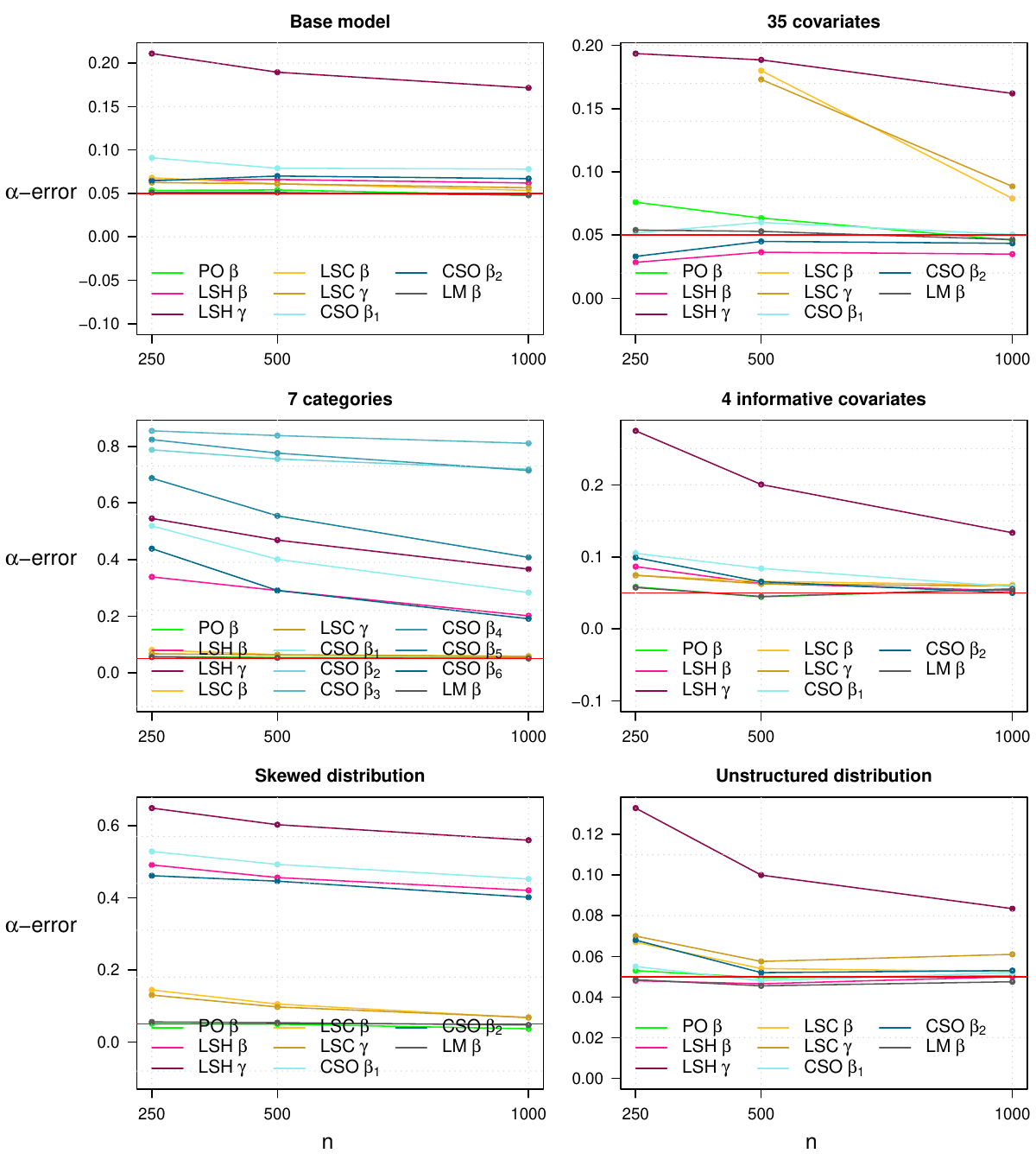}
    \caption{Influence of simulation settings on the $\alpha$-errors of an uninformative covariate of the model. Data was drawn from the location-shift model with $\beta = 1$ and $\gamma = 1$. The chosen base model has $p = 5$ covariates, $k = 3$ categories, 1 informative covariate, and the uniform distribution setting.}
    \label{alpha_disp_shift_v5_g1_b1}
\end{figure}

\begin{figure}[h]
    \centering
    \includegraphics[width=1\textwidth]{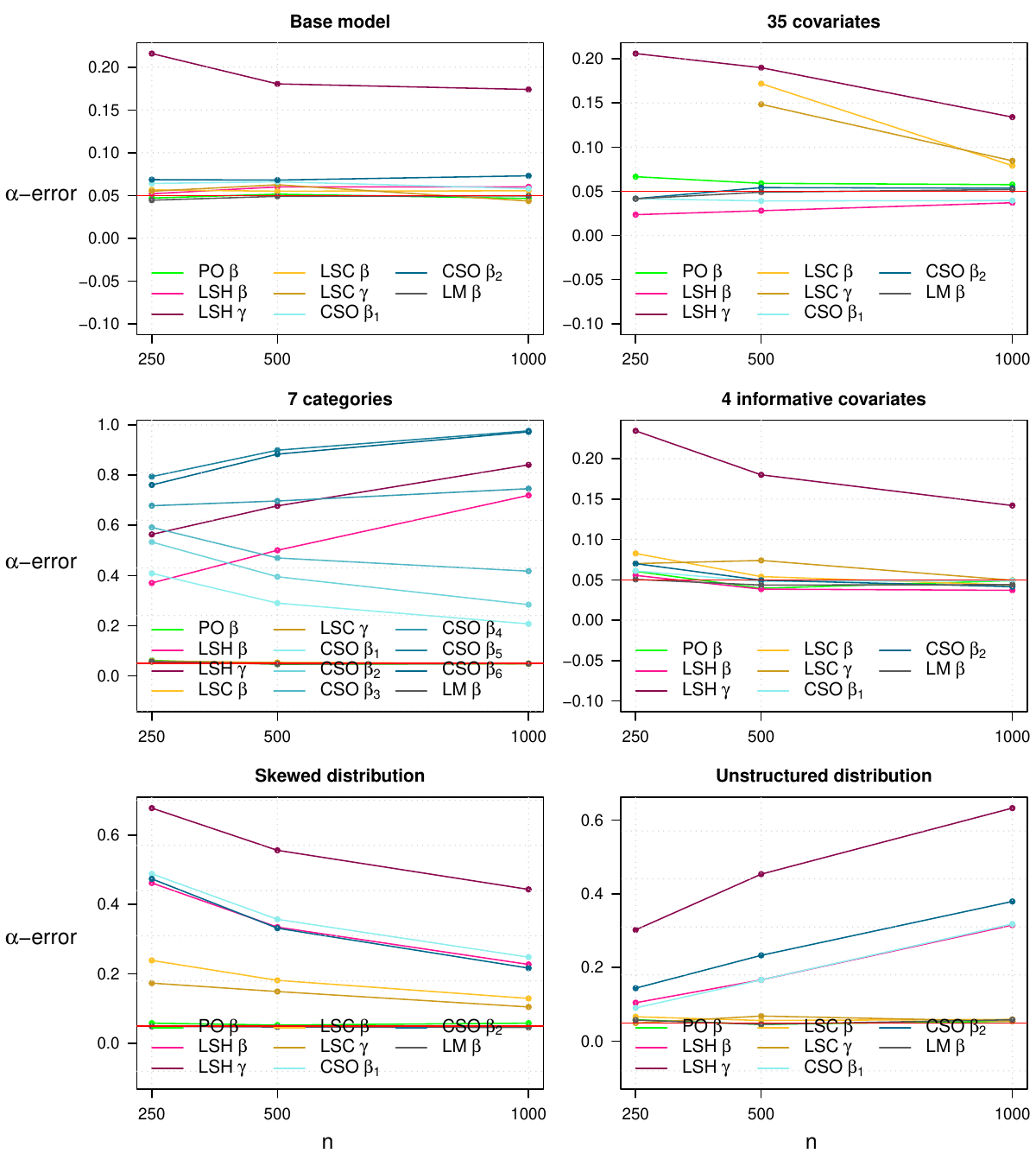}
    \caption{Influence of simulation settings on the $\alpha$-errors of an uninformative covariate of the model. Data was drawn from the location-scale model with $\beta = 1$ and $\gamma = 1$. The chosen base model has $p = 5$ covariates, $k = 3$ categories, 1 informative covariate, and the uniform distribution setting.}
    \label{alpha_disp_scale_v5_g1_b1}
\end{figure}

\paragraph{$\alpha$-Error when drawing from models with dispersion effect}
Figures~\ref{alpha_disp_shift_v5_g1_b1}~and~\ref{alpha_disp_scale_v5_g1_b1} display the $\alpha$-errors for the data generation process of the location-shift and the location-scale model, respectively. True parameters were chosen as $\gamma = 1$ and $\beta = 1$ for all informative covariates. To compare all models, we looked at an uninformative covariate. The setting $k = 5$ for the number of categories was mostly between the base setting with $k = 3$ and the setting for $k = 7$, so these results will not be displayed.

In comparison to data generation with the PO model, we saw higher $\alpha$-errors for the location-shift and the category-specific odds parameter tests across all settings.\newline 
In the base setting, the $\alpha$-error of the dispersion parameter test of the location-shift model was highly inflated with values around 0.2. The two models that were the closest to keeping the $\alpha$-level are the linear model and the proportional odds model. \newline
Increasing the number of covariates lead to more inflated $\alpha$-errors for both parameters of the location-scale model. The dispersion parameter of the location shift model stayed inflated, while the location parameter got more conservative. For small $n$, the proportional odds model showed slightly inflated $\alpha$-errors. However, for increasing $n$, this effect disappeared. The linear model was closest to the $\alpha$-level overall. Increasing the number of categories lead to strongly inflated $\alpha$-errors for all parameters in the category-specific odds model and the location-shift model. When drawing with the location-scale model, the inflated error rates seemed to increase with increasing $n$, especially for the location-shift model parameters. In the settings with 4 informative covariates instead of 1, the dispersion parameter of the location-shift model consistently showed the highest $\alpha$-error again. When drawing with the location-scale model, both parameters showed increased $\alpha$-errors as well, but only for $n \in \{250, 500\}$, while drawing with the location-shift model lead to specific increases of the $\alpha$-error of the parameters of the category-specific odds model for small $n$. Changing the distribution from uniform to skewed lead to highly inflated $\alpha$-errors for all parameters in the location-shift and the category-specific odds model. Similarly, but less affected, were the parameters of the location-scale model. The only two models close to keeping the $\alpha$-level are the linear model and the proportional odds model. These two models were also closest to keeping the $\alpha$-error for the unstructured distribution. Drawing with the unstructured distribution from the location-shift model lead to comparatively low $\alpha$-error inflation for the location-scale and category-specific odds model parameters and also for the location parameter of the location-shift model. The dispersion parameter still showed consistently high error rates. Using the unstructured distribution in the location-scale model lead to inflated $\alpha$-errors for all parameters of the location-shift and the category-specific odds model, which even increased with increasing $n$. The location-scale, proportional odds, and linear model parameters were all close to keeping the $\alpha$-level.

\clearpage

Further results for data generation with a dispersion parameter of $\gamma = 1$ and a location parameter of $\beta = 0$ can be found in Appendix~\ref{Further_res_alpha}. 

Figures \ref{alpha_disp_shift_v5_g1_b0} and \ref{alpha_disp_scale_v5_g1_b0} show the $\alpha$-errors of an uninformative variable for all models when drawing with the location-shift model and the location-scale model, respectively. Overall, we saw the proportional odds model and the linear model keep closest to the $\alpha$-error.

Figures~\ref{alpha_disp_shift_v1}~and~\ref{alpha_disp_scale_v1} show the $\alpha$-error for the location parameter of an informative variable where the dispersion parameter $\gamma = 1$ and the location parameter $\beta = 0$ when data was drawn from the location-shift model or the location-scale model, respectively.

In the location-shift setting, the linear model consistently upheld the $\alpha$-level for the uniform distribution, while the proportional odds model was close to the $\alpha$-level. For many covariates, the location parameter of the location shift model also kept the $\alpha$-level. When changing the distribution to the skewed case, all location parameters showed inflated $\alpha$-errors which did not improve with increasing $n$, with the linear model being closest to the $\alpha$-level. In the case of the unstructured distribution, only the location-shift model kept the $\alpha$-level. \newline
The location parameters of the location-scale, the proportional odds, and the linear model generally were at least close to keeping the $\alpha$-error overall when drawing with the location-scale model. The location shift model only upheld the $\alpha$-level for 35 covariates. In this case and for the skewed distribution, we saw the location parameter of the location-scale model display higher $\alpha$-error inflation. However, for the skewed distribution, all other models showed much higher $\alpha$-errors of about 1. In the case of the unstructured distribution, only the location-scale parameter upheld the $\alpha$-level, while the proportional odds model showed slightly increased $\alpha$-errors, the linear model showed increased $\alpha$-errors, and the location shift model showed highly increased $\alpha$-errors.

\newpage

\begin{figure}[h!]
    \centering
    \includegraphics[width=\textwidth]{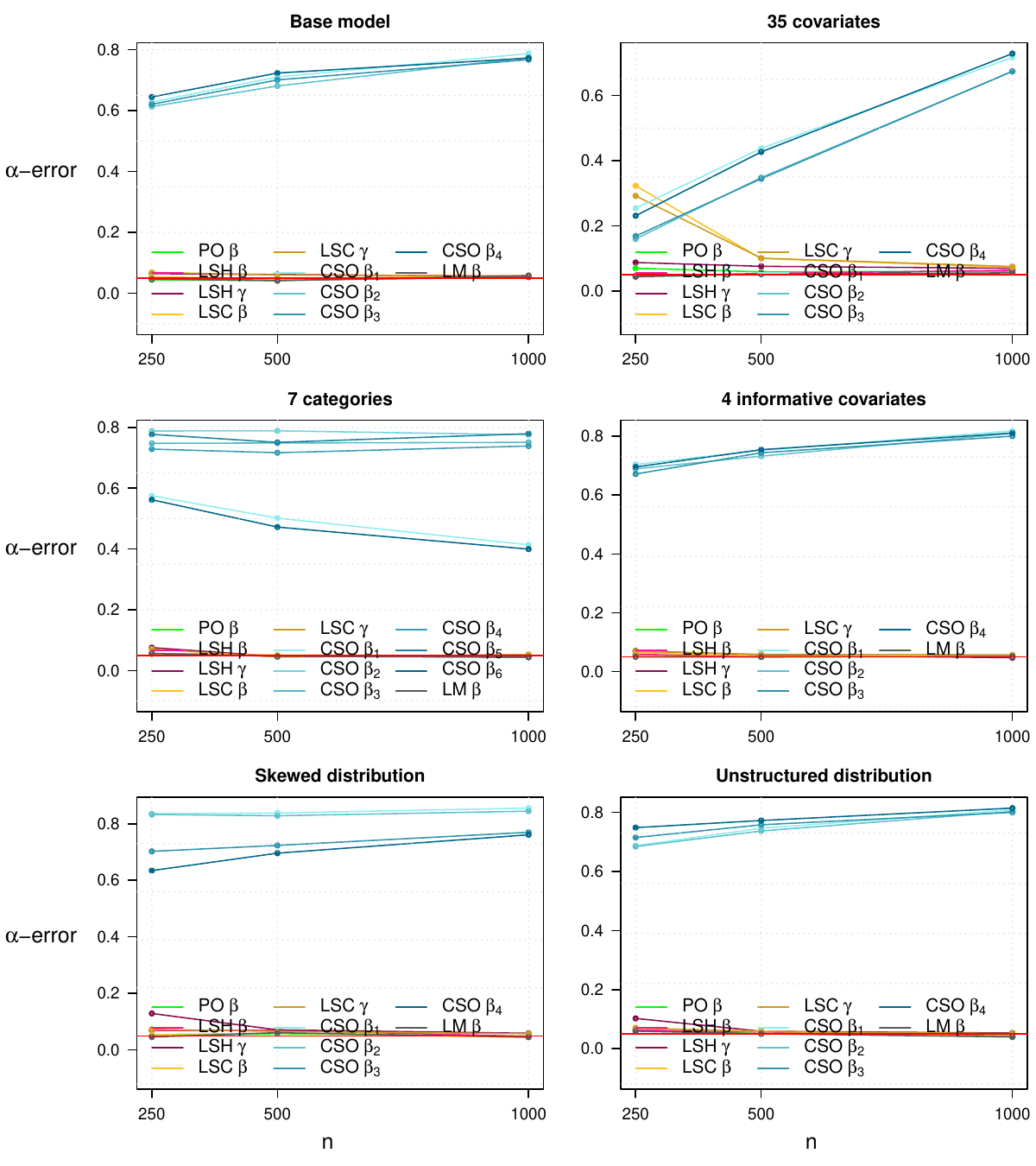}
    \caption{$\alpha$-error of the models when drawing with the CSO model. The chosen base model has $p = 5$ covariates, one informative covariate with $u = 1$, $k = 5$ categories, and the uniform distribution setting.}
    \label{alpha_cso_v5}
\end{figure}

\paragraph{$\alpha$-Error when drawing from the category-specific model}
Figure \ref{alpha_cso_v5} displays the $\alpha$-error for the data generation process with the CSO model with $u = 1$ and $\gamma = 0$ for all informative covariates. \mbox{$k = 5$} categories were chosen as baseline, because then the CSO model should profit from having more parameters~(4) in comparison to the dispersion models with two parameters.

In all settings with data drawn from the CSO model, the $\alpha$-error for the parameter tests in the CSO model was very highly inflated, and with increasing $n$, the errors increased overall. For 35 covariates, the location-scale model showed highly inflated $\alpha$-errors for both parameters for $n = 250$. However, with increasing $n$, these errors decreased. The location-shift model showed inflated $\alpha$-errors in most settings with $n = 250$ observations, which decreased with increasing $n$. The closest to keeping the $\alpha$-level were the linear model and the proportional odds model.

Figure~\ref{alpha_cso_vs_slight_vars} in Appendix~\ref{Further_res_alpha} displays results for the $\alpha$-error of the category-specific odds model parameters for data drawn from the CSO model. We present results for an informative covariate with $u = 1$, so that in the case of 7 covariates, $\beta_2 = \beta_6 = 0$. We observed that for every parameter setting, the $\alpha$-errors for $\beta_2 \text{ and }\beta_6$ were highly inflated, with $\alpha$-errors of 0.6 at the lowest.

\clearpage

\subsection{Power}
The power evaluation is performed for hypothesis tests for the relevant parameter to be zero, where the true parameter takes all predefined values. At the point of the power curve where the real parameter value is 0, the curve displays the $\alpha$-error. All informative covariates had the same true parameter values.

\begin{figure}[h!]
    \centering
    \includegraphics[width=1\textwidth]{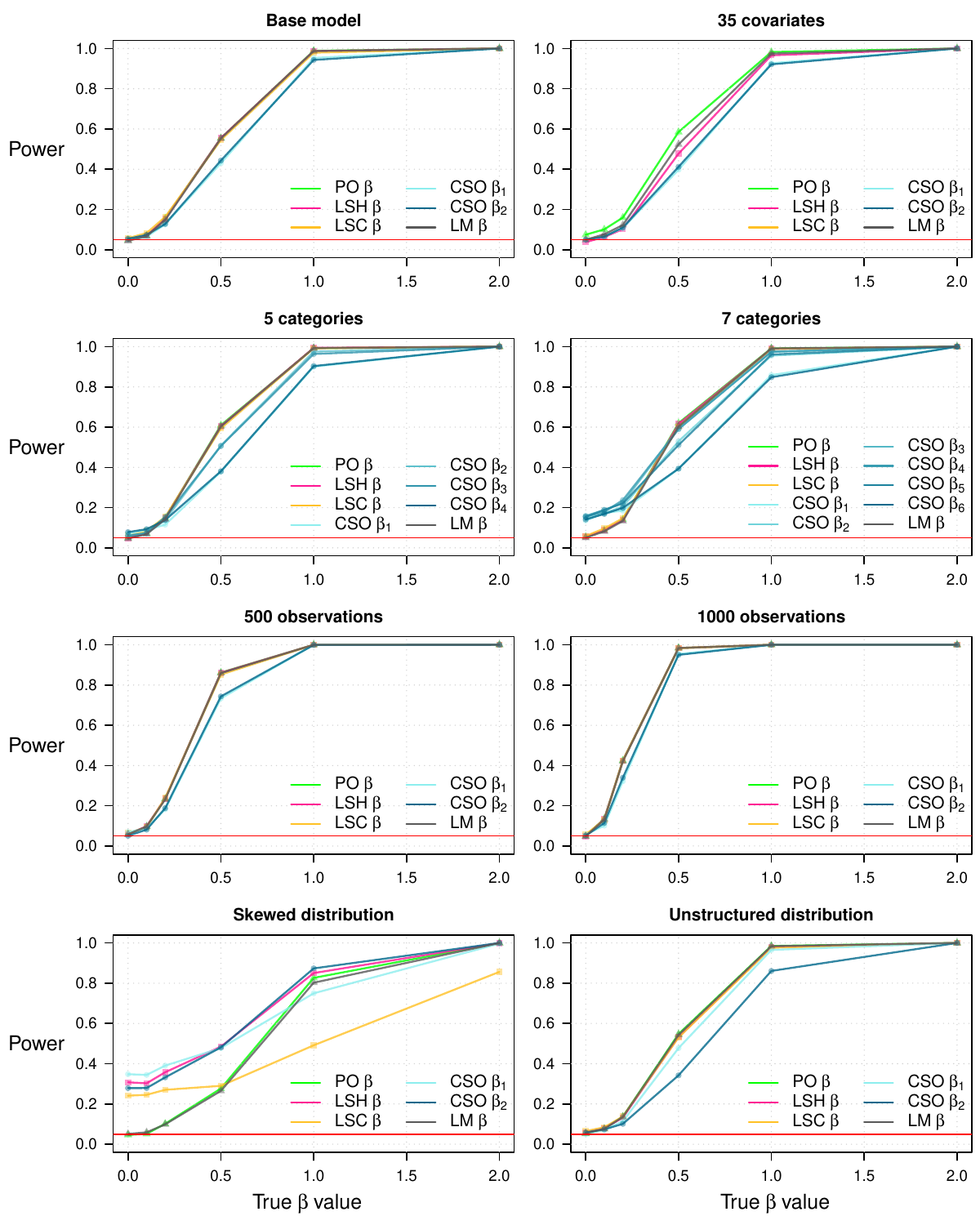}
    \caption{Power of models comparatively to the chosen base model with $p = 5$ covariates, one of them informative, $k = 3$ categories, 250 observations and the uniform $\theta$ setting. Data was generated with a proportional odds model. For $\beta = 0$ the plots show the $\alpha$-error.}
    \label{power_po}
\end{figure}

\subparagraph{Power of the location parameters when drawing from the proportional odds model}
In the following, we chose the simulation setting with $p = 5$ covariates, of which one was informative, $k = 3$ categories, $n = 250$ observations, the uniform distribution, and data drawn from the proportional odds model with varying $\beta$ values as the baseline model. Figure~\ref{power_po} displays the power curve for this base setting and for variations where one parameter was changed. 

In the baseline settings for $\beta = 0$, most models started close to the $\alpha$-level and increased their power with increasing $\beta$. The proportional odds model, the location parameter of the location-shift and the location-scale model, and the linear model all showed slightly higher power for medium-sized $\beta$-values than the tests for the parameters of the category-specific odds model.

Since the number of informative covariates had the least impact on the power curve, it is not considered further.\newline
Increasing the number of covariates to 35 did not meaningfully alter the power curves, although the tests overall are slightly more conservative. Location-scale models mostly did not converge and could not be considered here. \newline
Increasing the number of categories only clearly influenced the category-specific odds model. The higher the number of categories, the higher the $\alpha$-error got and the flatter the power curve got as well. \newline
As the number of observations increased,  all power curves became steeper. However, the tests for the parameters of the category-specific odds model stayed consistently equal to or performed worse than those of the other models.\newline
Changing the distribution from uniform to skewed lead to strong changes: The location parameter of the location-scale model showed a very flat power curve, and all parameters in the location-shift, the location scale, and the category-specific odds model showed highly inflated $\alpha$-errors. Only the linear model and the proportional odds model showed acceptable $\alpha$-errors and still reached similar power to the other models with relatively high $\beta$ values. For the unstructured distribution, the power curve was similar to the uniform setting.

\clearpage

\newpage

\begin{figure}[h!]
    \centering
    \includegraphics[width=1\textwidth]{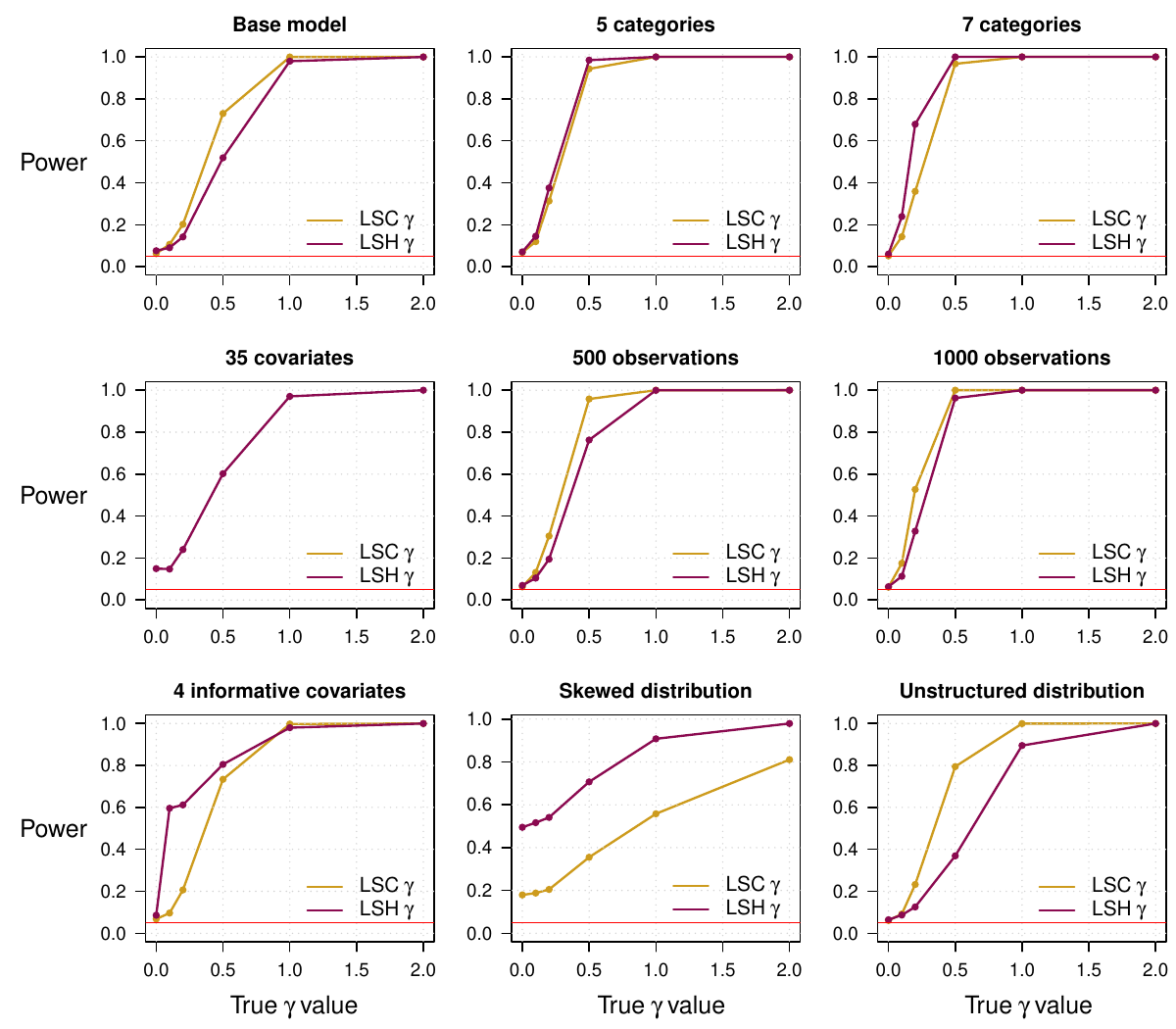}
    \caption{Power of models with dispersion estimations compared to the chosen base model with $p = 5$ covariates, one of them informative, $k = 3$ categories, 250 observations, uniform $\theta$ setting and $\beta = 0$. The dispersion parameter of the location-scale model was calculated with the data generation process of the location-scale model, and the dispersion parameter of the location-shift model was calculated with the location-shift model. For $\gamma = 0$, the plots display the $\alpha$-error.}
    \label{power_disp}
\end{figure}

\subparagraph{Power of dispersion parameters when drawing from models with dispersion effect}
Figure~\ref{power_disp} shows the power curve for the tests for the dispersion parameters of the location-shift and the location-scale model. To ensure we know the true dispersion value, the location-shift models were calculated on data generated with the location-shift model, and the location-scale models were calculated on data generated with the location-scale model. The simulation settings chosen as base model have $p = 5$ covariates, one of them informative, $k = 3$ categories, 250 observations, uniform $\theta$ setting, and $\beta = 0$.

In the base setting, we observed that the tests for the dispersion parameters almost kept the $\alpha$-level and increased with increasing true $\gamma$. The power curve of the dispersion parameter of the location-scale model was a little steeper. \newline
We also considered settings withK$\beta = 1$, but the results were very similar to the base model and are omitted for brevity. \newline
When increasing the number of categories, the power curve for both models got steeper. The dispersion parameter of the location-shift model benefited more from a higher number of categories and had higher power than the location-scale model for 7 categories with small to moderate $\gamma$ values. \newline
When looking at the base settings, but with 35 covariates, we saw that most of the location-scale models did not converge. The dispersion parameter of the location-shift model did not adhere to the $\alpha$-level. \newline
Increasing the number of observations lead to steeper power curves, while keeping the test for the dispersion parameter of the location-scale model as the slightly more powerful test.\newline
When choosing 4 instead of one covariate, we saw that the location-shift model outperformed the location-scale model for low values of $\gamma$. \newline
Changing the distribution from uniform to skewed resulted in highly inflated $\alpha$-errors, especially for the location-shift model and relatively flat power curves. Again, the setting with an unstructured distribution looked relatively similar to the uniform one. However, the power was slightly lower for the tests of the dispersion parameter in the location-shift model.

\clearpage

\newpage

\begin{figure}[h!]
    \centering
    \includegraphics[width=\textwidth]{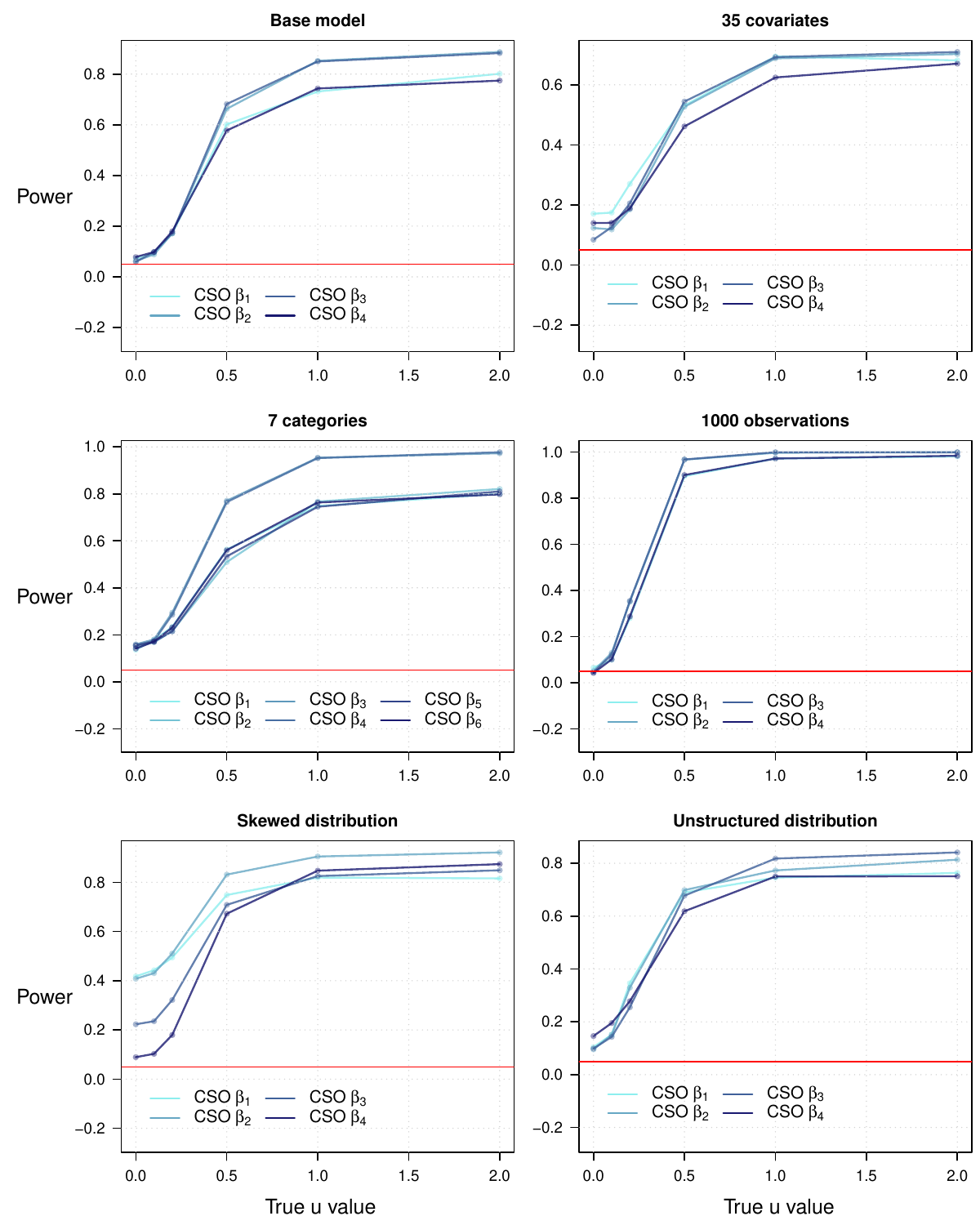}
    \caption{Power of the CSO parameter tests when drawing with the CSO model compared to the chosen base model with $p = 5$ covariates, one of them informative, $k = 5$ categories, 250 observations, and uniform $\theta$ setting. For seven categories, the true $\beta_2$ and $\beta_5$ were chosen to be zero. In the case where the true value of $u$ was zero, the plots display the $\alpha$-error for all models.}
    \label{power_cso}
\end{figure}

\subparagraph{Power when drawing from the category-specific model}
Figure~\ref{power_cso} shows the power curve for the drawing process with the category-specific odds model. For five categories this means the true parameter vector $\boldsymbol{\beta} = (-u, u, u, -u)$ and for seven categories $\boldsymbol{\beta} = (-u, 0, u, u, 0, -u)$. On the x-axis, we always displayed the value of $u$. As the base setting, we chose $p = 5$ covariates, one of them informative, $k = 5$ categories, 250 observations, and the uniform $\theta$ distribution. 

For the category-specific odds model, we saw that $\beta_2$ and $\beta_3$ (positive parameter value) had slightly steeper power curves and a slightly lower $\alpha$-error than $\beta_1$ and $\beta_4$ (negative parameter value). \newline
The number of informative covariates had the least relevant influence on the results and is therefore not displayed.\newline
Looking at the results for 35 covariates, the $\alpha$-errors were higher than for five categories, especially for $\beta_1$. The power for these parameters was still at maximum 0.7 for a $u$ value of 2.\newline
For seven categories, the true regression coefficient vector for the data generation was $\boldsymbol{\beta} = (-u, 0, u, u, 0, -u)$, which means for these results, Figure~\ref{power_cso} displays the $\alpha$-error for $\beta_2$ and $\beta_5$. However, this could not be recognized in the figure, as with increasing $u$, the $\alpha$-error was getting increasingly inflated for $\beta_2$ and $\beta_5$ until about 0.8 for $u = 2$, which was on par with $\beta_1$ and $\beta_6$. The power curve for the tests for $\beta_3$ and $\beta_4$ were steeper, almost reaching 1 for $u = 2$, similar to the base model.\newline
For 1000 observations, the $\alpha$-level for $u = 0$ was roughly held by all parameters, and we saw a steep power curve for the parameters of the CSO.\newline
When changing the distribution to the skewed setting, $\beta_1$, $\beta_2$, and $\beta_3$ of the category-specific odds model showed highly inflated $\alpha$-errors, and $\beta_4$ a slightly inflated $\alpha$-error. The unstructured distribution setting appeared similar to the uniform setting.

\clearpage

\newpage

\newpage

\section{Discussion}
In this paper, a simulation study was conducted to examine the properties of the parameter estimates of the proportional odds model (POM), the category-specific odds model (CSO), the location-shift model (LSH), and the location-scale model (LSC). The linear model (LM) was included, given the ongoing discussion on whether normal approximations for ordinal data are acceptable in practical settings \citep{kniefViolatingNormalityAssumption2021}, or whether more complex models should be used \citep{liddellAnalyzingOrdinalData2018}. 

Performing the simulations, we observed that the location-scale model exhibited more convergence issues than all other investigated models, particularly for many covariates, few observations, and a low number of categories.

Firstly, we examined the bias of the parameter estimations of the POM, the CSO, the LSH, and the LSC. Regarding biases, the CSO and the LSH tended to underestimate the true parameters, whereas the LSC and the POM tended to overestimate them. For the location parameters across all models, we observed especially high biases for high values of the true parameter. For dispersion parameters, high values lead to strong biases for the location-shift and smaller biases for the location-scale models. Biases of the location and the dispersion parameters of the location-shift model were positively correlated. The number of categories and informative variables primarily increased the negative bias in the dispersion parameter of the location-shift model, and the location parameter of the category-specific model slightly, but did not affect the bias of the other models.  

Afterwards, we investigated the $\alpha$-error for the parameter tests of the POM, the CSO, the LSH, the LSC, and the LM. When generating data from the POM, the linear model and the proportional odds model were closest to keeping the intended $\alpha$-level. Especially when a skewed distribution was chosen, all other models showed highly inflated errors. During data generation from the models with dispersion parameters, again, the linear model and the proportional odds model were the best at keeping the $\alpha$-level. Only for 35 covariates, the $\alpha$-error of the POM was inflated. For the data generation process with the category-specific odds model, we observed especially high inflated $\alpha$-errors for the CSO across all settings, while the linear model and the proportional odds model kept relatively close to the $\alpha$-level.

Similarly, we reported the respective power curves, where again, the linear model and the POM appeared to be best suited for different data generation settings. Especially when a skewed distribution was chosen, the power curve for the parameter tests for both parameters of the location-scale model and the dispersion parameter of the location-shift model was very flat. \newline
\cite{liddellAnalyzingOrdinalData2018} suggested, when the mean of observations is close to the end of a scale, metric models may produce errors. The skewed distribution was chosen to examine this case, and at least for the linear model in our simulation study, the gravity of the errors was small in comparison to most other models examined.

In conclusion, among the examined models, the linear model and the proportional odds model demonstrated the best overall inferential properties in our simulation study. They kept the $\alpha$-level across almost all simulation settings and showed comparatively steep power curves when drawing with the proportional odds model. Among ordinal models, the strong performance of the proportional odds model is in line with earlier findings \citep{buczakOldGoldNew2024,tutzOrdinalRegressionReview2022}. The good performance of the linear model supports the claim that the linear model is rather robust against violations of the normality assumption \citep{kniefViolatingNormalityAssumption2021} when sample sizes are rather large, as in our study ($n\geq 250$). However, its interpretation is more challenging since ordinal categories cannot be treated as equidistant data points. Based on our results, we thus generally recommend the use of simpler regression models, particularly the proportional odds model, over more complex alternatives, when in doubt. Models with a dispersion parameter should only be used if there are very relevant contraindications to the use of just one location parameter, and the inferential properties of the CSO lead us to advise against this model in general.

\subsection{Limitations}
In this simulation study, the influence of all informative covariates on the target variable was always simulated with identical regression weights. This ensures that with increasing regression parameter values, the influence of all informative covariates increases similarly and influences on the properties of the model are not due to different values in regression parameters across covariates but due to the value of the regression parameter itself. Nonetheless, the argument can be made that varying the regression parameter values in the settings with four informative covariates should only influence the parameters of the variable of interest and may lead to different results in the power analysis. 

The data generation processes were chosen with the covariates being normally distributed with a mean of zero and a standard deviation of 0.5 to limit the number of cases with negative probabilities. Furthermore, if negative probabilities appeared, the observations were discarded. In reality, this data exists and is neglected in our simulation study. Different distributions and a sensible way to deal with negative probabilities may be needed in future simulations.

As there are 4032 simulation settings and 2000 simulation repetitions, there were many models that showed warnings in R. In this simulation study, these warnings were ignored. In application scenarios, we would naturally advise against ignoring such warnings, as they may indicate issues that impact result reliability. This may have lead to worse performances of the more complex models, like the category-specific, the location-shift, and the location-scale model.

In this simulation study, we focus on inferential properties. Nonetheless, there are various further ways to evaluate the performance of a model. Future studies might aim to investigate the predictions of these models. This also allows us to extend comparisons with other ordinal regression models like tree-based models \citep{hornungOrdinalForests2020,buczakOldGoldNew2024,buczak2025frequency}.

Finally, we used Wald-type tests for our simulation. These tests are simple and are used in many packages \citep[e.g.][]{yeeVGAMPackageCategorical2010}. However, they may not be ideally suited for ordinal outcomes and work more reliably with a larger number of observations. These drawbacks may be avoided by using resampling methods, as proposed for nonparametric analyses in factorial designs \cite[e.g.][]{umlauft2017rank,dobler2020nonparametric}. 
\newpage

\subsection*{Declarations}
\paragraph{Funding} No funding was received for conducting this study.
\paragraph{Conflicts of interest} The authors have no competing interests to declare that are relevant to the content of this article.
\paragraph{Ethics approval} Not applicable.
\paragraph{Consent to participate} Not applicable. 
\paragraph{Consent for publication} Not applicable.
\paragraph{Availability of data and materials} This study only uses artificially generated (simulation) data and no personal data. 
\paragraph{Code availability} The code and all simulated data sets will be published as supplementary material.

\subsection*{Open Practices}
The code and all simulated data sets will be published as supplementary material. The study itself was not preregistered.
\bibliographystyle{apacite}
\bibliography{OrdinalRegression_bib}

\newpage

\appendix
\section{Literature review details}

To gain a clearer understanding of which models are used for ordinal outcomes in practice, we conducted a literature review using PubPsych (https://pubpsych.zpid.de/pubpsych/). We used the search terms "regression" and either "ordinal" or "ordered" and only included studies that were published between 2014 and 2023.\newline

We obtained 676 articles, of which 16 were not accessible. After excluding all papers that did not explicitly report the use of an ordinal model, we were left with 458 papers. Among these, 6 papers used two different methods, and one paper used 4 different methods; the specific models can be found in Table~\ref{lit_rev}.

We classified a paper as using a specific model, when either the paper explicitly named the model, stated the respective formulas, listed the assumptions, mentioned the model as having passed respective assumption tests or if the model was calculated with the logit link in SPSS without extra packages, as to our knowledge, there is only an implementation of the proportional odds model and no other ordinal regression model.

In 11 papers, it was unclear how or where an ordinal model was used, or no details about the model were given. 50 papers implemented some form of mixed model, and 4 used Bayesian methods, which both extend beyond the scope of this study. 

The models used in this simulation study that were named in the literature review were the proportional odds model appearing in 205 papers (52\%), the partial proportional odds model in 14 papers (3.54\%), and the category-specific odds model in 2 papers (0.5\%). One paper reported a location and a scale parameter. Other papers used different link functions than the logit link: We observed 12 papers with a probit link, 5 with a negative log-log link and one with a log-log link. 

136 papers used the logit link function but did not explicitly state which model was used. Out of these papers, 105 papers reported one odds ratio per covariate, 14 reported one regression weight per covariate, and 7 reported both, while 10 reported none. 23 papers did not state the link function explicitly, with 1 paper reporting an odds ratio and one regression parameter per covariate, 5 reporting no odds ratios, but one regression parameter per covariate, and 17 papers reporting an odds ratio but no regression parameters. A detailed listing can be seen in Table \ref{lit_rev}. \newline

\begin{table}[htpb]
\caption{Literature review results}
\label{lit_rev}
\begin{tabular}{cccc}
\toprule
Type of model & Number & Percentage & Percentage \\
& of studies & respective to  & respective to\\
& & Total (1) &  Total (3) \\
\midrule
Mixed &  47 & .1006 & \\ 
        Bayes &  2 &  .0043 &   \\ 
        Bayes Mixed &  2 &  .0043 &  \\ 
        Unclear &  11 &  .0236 &  \\ 
        Other &  6 &  .0128 &  .0152 \\ 
        Probit Link &  11 &  .0236 &  .0278 \\ 
        Probit mixed &  1 &  .0021 &  .0025 \\ 
        Loglog link &  1 &  .0021 &  .0025 \\ 
        Negative Loglog Link &  5 &  .0107 &  .0126 \\ 
        Proportional Odds (PO)	&205&	.4390&	.5177\\
        Partial Proportional Odds (PPO) &14&	.0300&	.0354\\
        Category-specific Odds (CSO) &2&	.0043&	.0051\\
        Location + scale parameter&1&	.0021&	.0025\\
        Log-link rep, no OR rep, no param rep & 10&	.0214&	.0253\\
        Log-link rep, OR rep, no param rep&105&	.2248&	.2652\\
        Log-link rep, OR rep, one param per covariate rep&7&	.0150&.0177\\
       Log-link rep, no OR rep, one param per covariate rep&14
       &.0300&.0354\\  
       No link rep, OR rep, one param per covariate rep&1
       &.0021&.0025\\  
       No link rep, no OR rep, one param per covariate rep&5
       &.0107&.0126\\  
       No link rep, OR rep, no param rep&17
       &.0364&.0429\\  
       \midrule
      Total different models in studies (1)&467&&\\  
      \midrule
      PPO and PO&2&&\\  
      Loglog link and location + scale parameter&1&&\\  
      Mixed and ordinal model with OR estimated &1&&\\  
      Probit link and mixed with probit link&1&&\\  
      PPO and CSO&1&&\\  
      PPO, PO,
     CSO and other model&1&&\\   
     \midrule
     Total different studies with ordinal models (2) &458&&\\
     \midrule
     Total excluding mixed and bayes and unclear (3)&396&&\\
\bottomrule
\end{tabular}

\bigskip
\small\textit{Note}. PO = proportional odds model; PPO = partial proportional odds model; CSO = Category-specific odds model; OR = Odds ratio; rep means reported;
\end{table}

\section{Problems in the simulation}
\label{problems_appendix}

In the data generation process, in case the condition $P(Y \leq r | \mathbf{x}) \leq P(Y \leq r+1 | \mathbf{x})$ did not hold, the data was redrawn. For our simulation study, this happened especially often in the case of drawing data from the category-specific model with high $\beta$ values and four informative covariates. In these cases, for one observation that held the condition, on average, about 4-6 observations had to be drawn. For all other settings, this number was lower. \newline

If the observed number of observations per category was less than 5, the whole dataset was redrawn. This happened especially often in the data generation setting with large dispersion, a low number of observations, a high number of categories, and the skewed distribution setting. In some of these cases, up to 1540 samples had to be redrawn. For all other settings, the number of redrawn samples was lower than 200. \newline

Among the cumulative models of interest, the category-specific odds model, the location-shift model, and the location-scale model showed relevant problems with convergence in some settings. The proportion of non-converging proportional odds models was less than $0.01\%$ in any setting.

\begin{figure}[h]
    \centering
    \begin{minipage}[b]{0.5\textwidth}
        \centering
        \includegraphics[page=1, width=\textwidth]{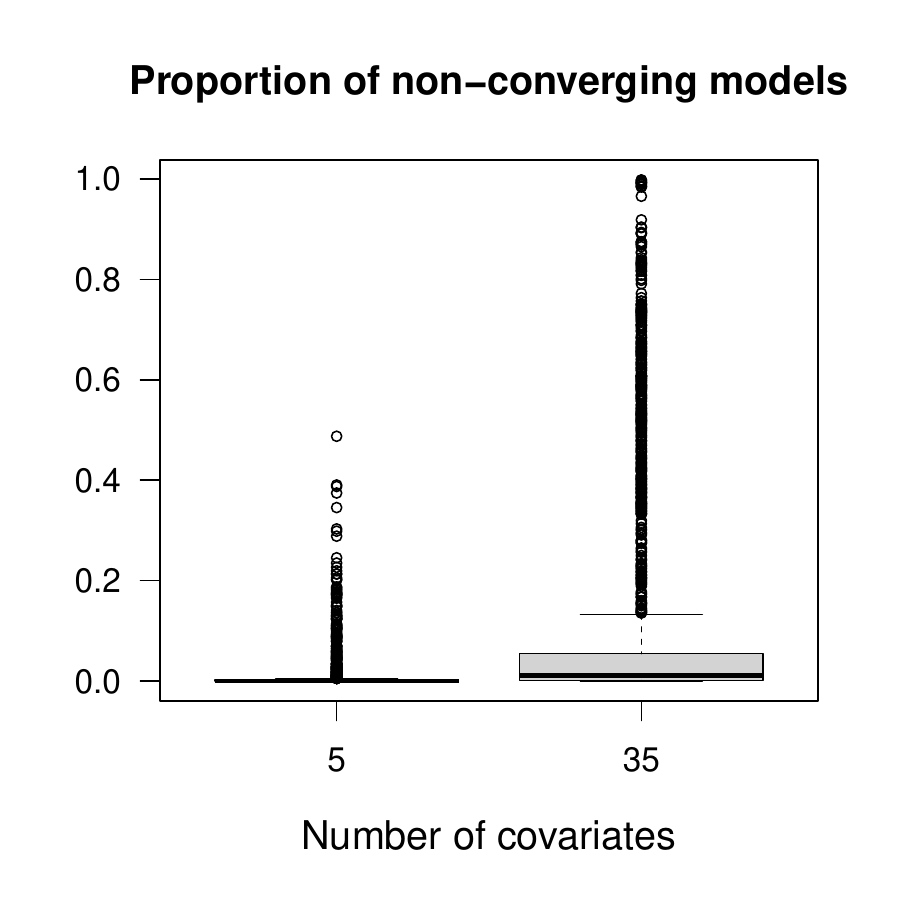}
    \end{minipage}
    \hspace{-1.2cm}
    \begin{minipage}[b]{0.5\textwidth}
        \centering
        \includegraphics[page=2, width=\textwidth]{figures/convergence_issues.pdf}
    \end{minipage}
    
    \begin{minipage}[b]{0.5\textwidth}
        \centering
        \includegraphics[page=3, width=\textwidth]{figures/convergence_issues.pdf}
    \end{minipage}
    \caption{Relevant influences on convergence of category-specific models.}
    \label{convergence_problems_cat_spec}
\end{figure}

\newpage
\subparagraph{Non-convergence in the category-specific model}
In 1537 out of the 4032 simulation settings, all category-specific models converged. In 382 settings, at least 20\% of the models did not converge. In Figure \ref{convergence_problems_cat_spec} the most influential simulation settings for non-convergence of models are displayed. Category-specific models have issues with convergence when the number of covariates is rather high, the number of observations is low, and the true $\gamma$ value is rather high.

\newpage

\subparagraph{Non-convergence in the location-shift model}

\begin{figure}[h]
    \centering
    \begin{minipage}[b]{0.5\textwidth}
        \centering
        \includegraphics[page=4, width=\textwidth]{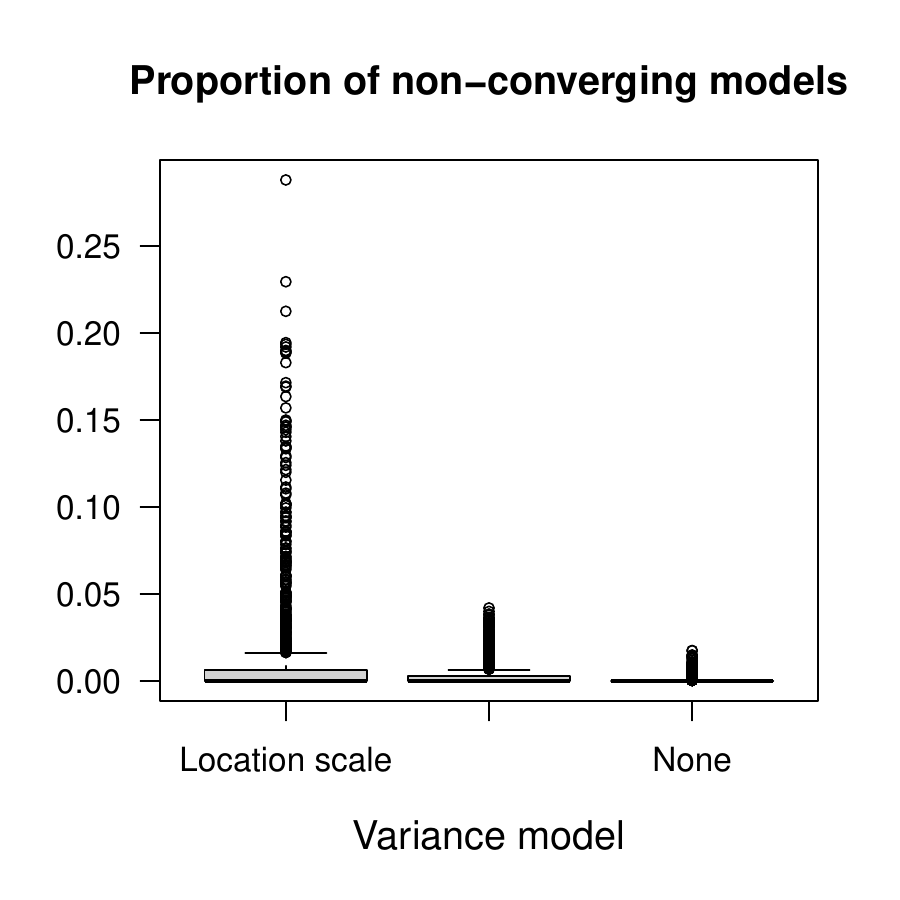}
    \end{minipage}
    \hspace{-1.2cm}
    \begin{minipage}[b]{0.5\textwidth}
        \centering
        \includegraphics[page=5, width=\textwidth]{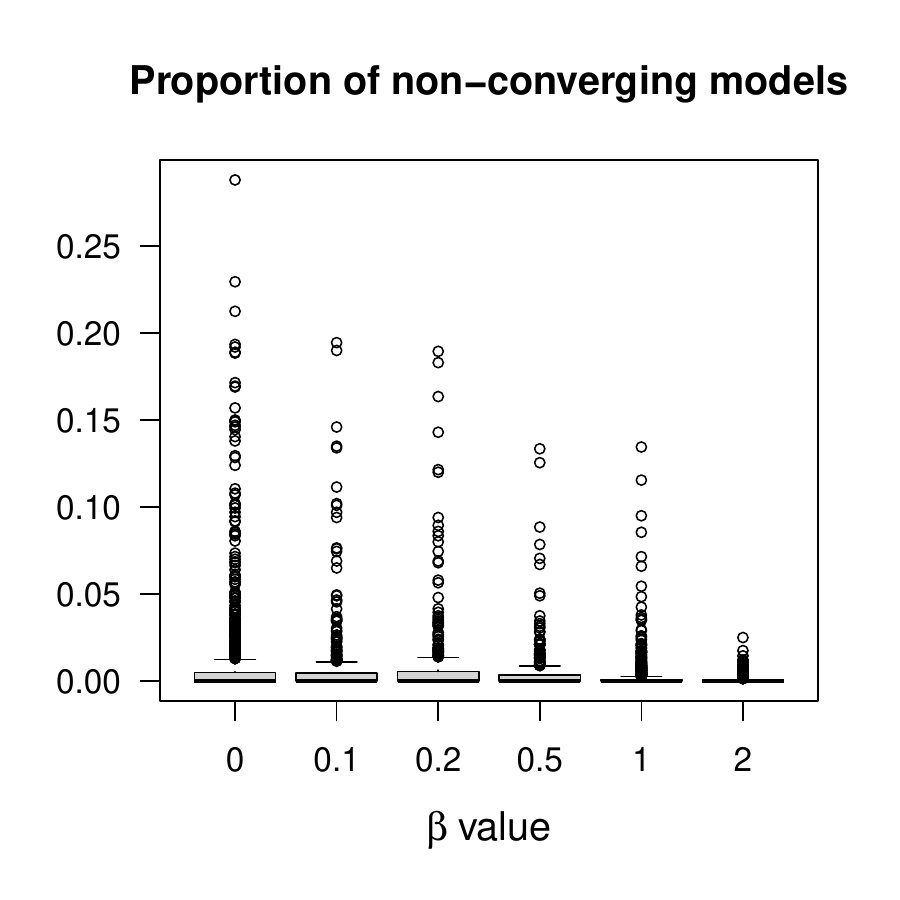}
    \end{minipage}
    
    \begin{minipage}[b]{0.5\textwidth}
        \centering
        \includegraphics[page=6, width=\textwidth]{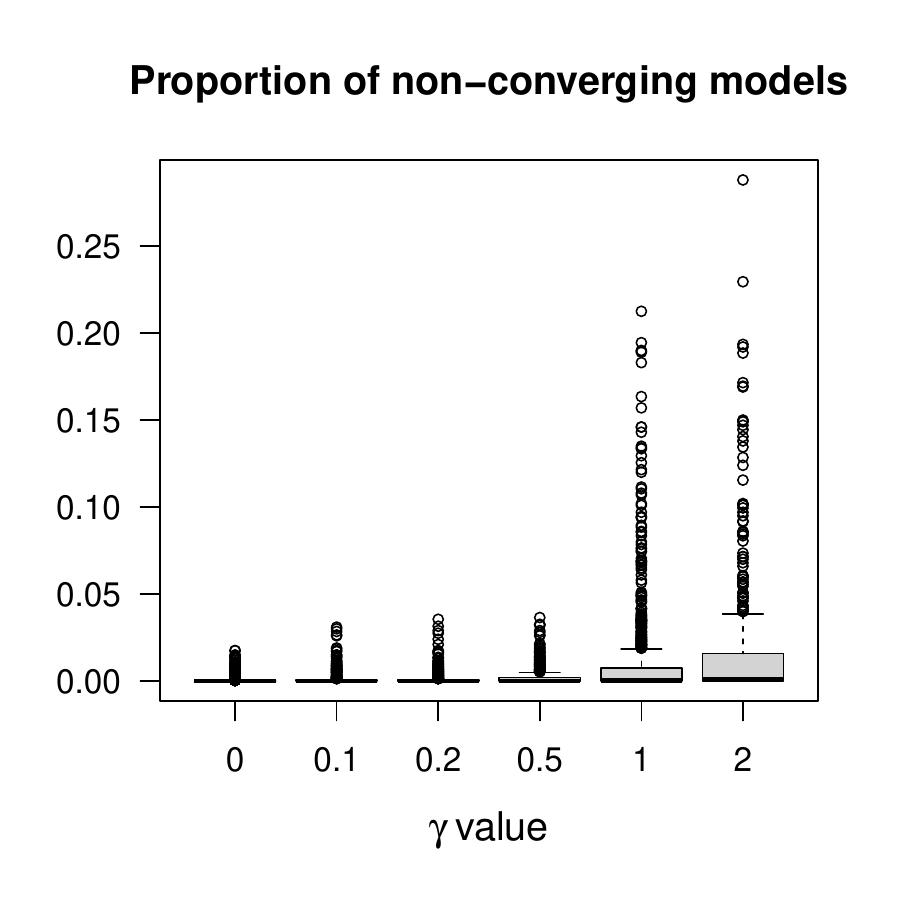}
    \end{minipage}
    \caption{Relevant influences on convergence of location-shift models.}
    \label{convergence_problems_loc_shift}
\end{figure}

In 2454 of the 4032 simulation settings, all location-shift models converged. In 3 settings, at least 20\% of the models did not converge. In Figure \ref{convergence_problems_loc_shift}, the most influential simulation settings for the non-convergence of models are displayed. Location-shift models had issues with convergence when the data generation process was performed with the location-scale model and higher $\gamma$ values, and there were rather low true $\beta$ values.

\newpage

\subparagraph{Non-convergence in the location-scale model}

\begin{figure}[h]
    \centering
    \begin{minipage}[b]{0.5\textwidth}
        \centering
        \includegraphics[page=7, width=\textwidth]{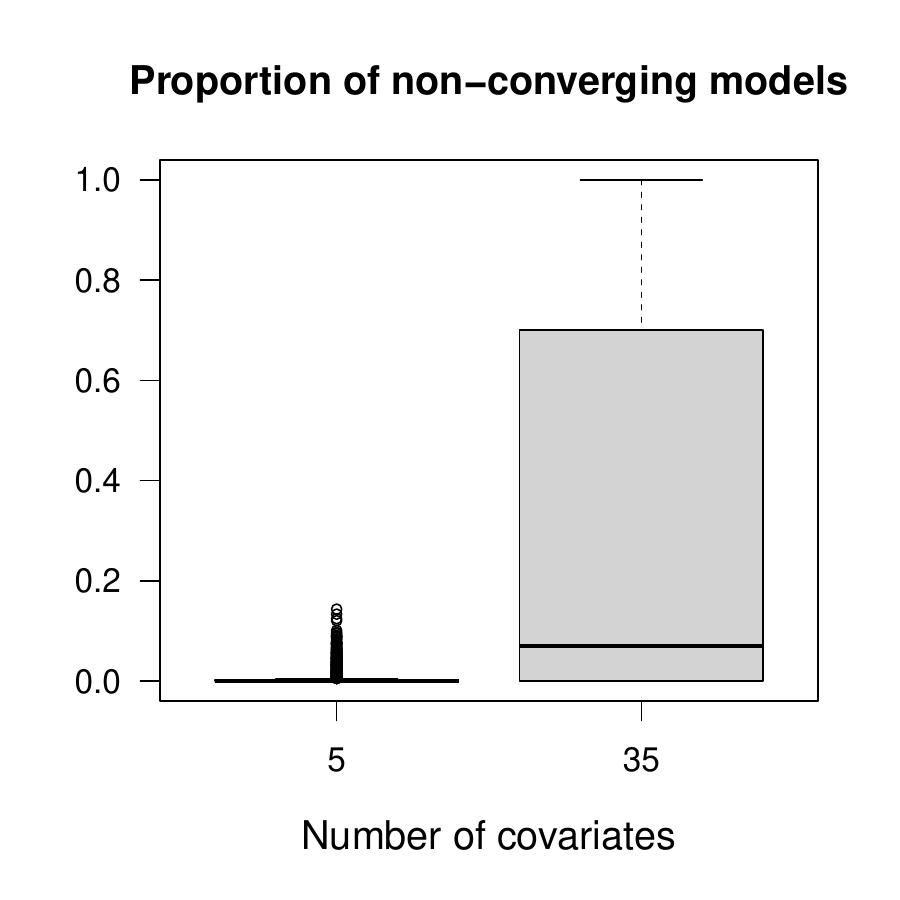}
    \end{minipage}
    \hspace{-1.2cm}
    \begin{minipage}[b]{0.5\textwidth}
        \centering
        \includegraphics[page=8, width=\textwidth]{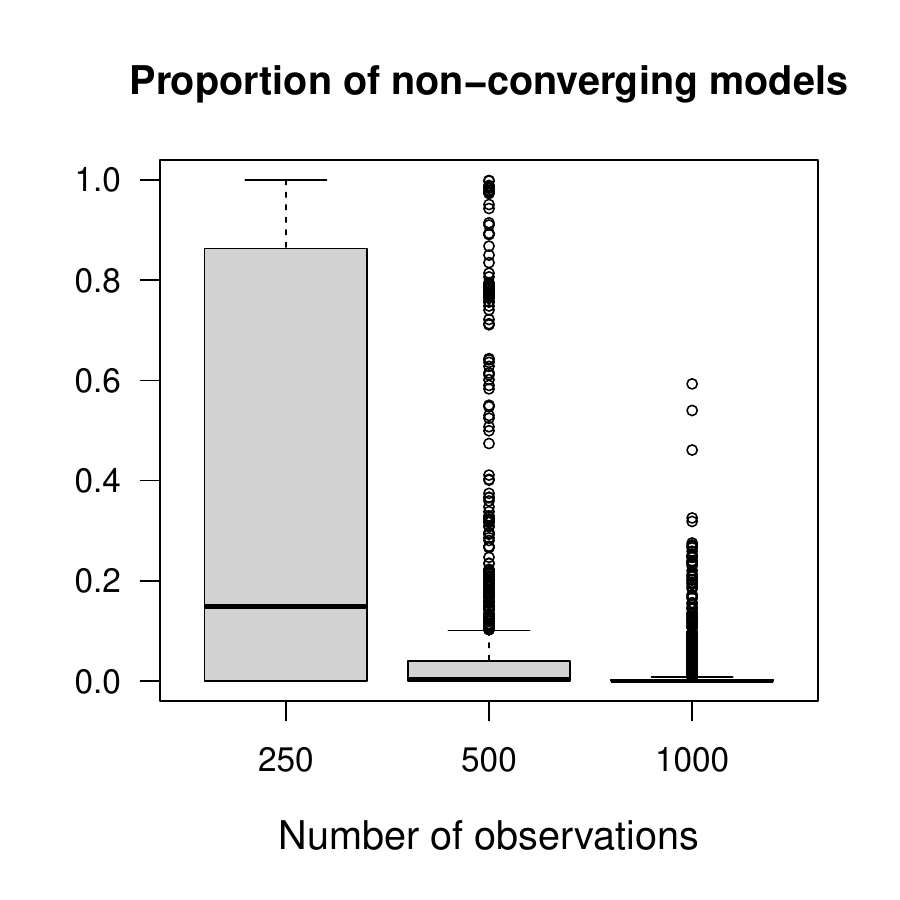}
    \end{minipage}
    
    \begin{minipage}[b]{0.5\textwidth}
        \centering
        \includegraphics[page=9, width=\textwidth]{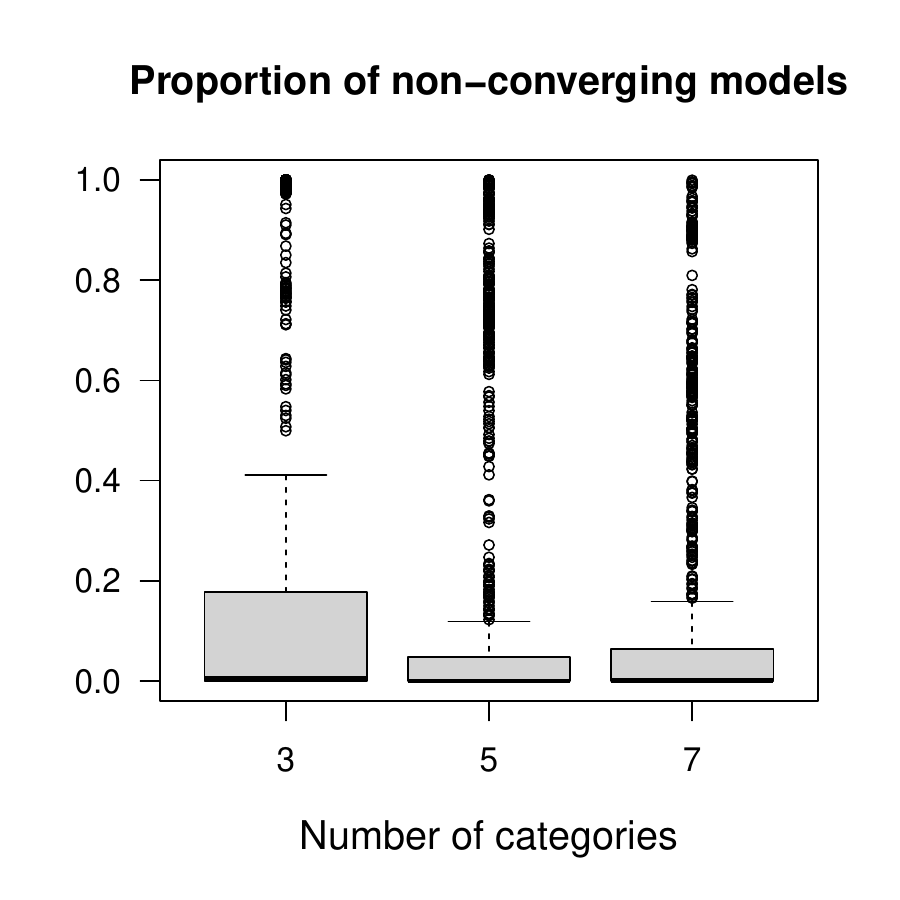}
    \end{minipage}
    \hspace{-1.2cm}
    \begin{minipage}[b]{0.5\textwidth}
        \centering
        \includegraphics[page=10, width=\textwidth]{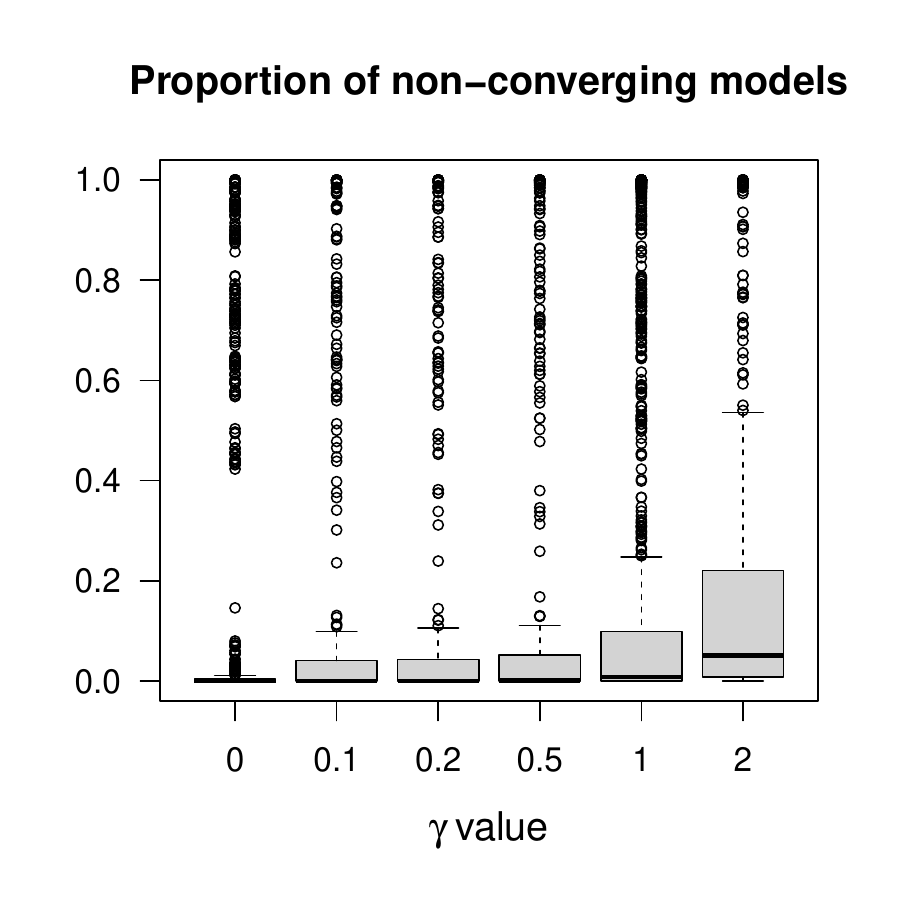}
    \end{minipage}
    \caption{Relevant influences on convergence of location-scale models.}
    \label{convergence_problems_loc_scale}
\end{figure}

In 1827 of the 4032 simulation settings, all location-scale models converged. In 807 settings, at least 20\% of the models did not converge, and for 66 simulation settings, no models converged. In Figure \ref{convergence_problems_loc_scale}, the most relevant simulation settings influencing the non-convergence of models are displayed. Location-scale models had issues with convergence when the data generation process was performed with the location-scale model and higher $\gamma$ values, and there were rather low $\beta$ values. 

\newpage

\section{Bias in the location parameter of the location-scale, location-shift and category-specific odds model}
\label{further_biases}

\begin{figure}[h]
    \centering
    \includegraphics[width=1\textwidth]{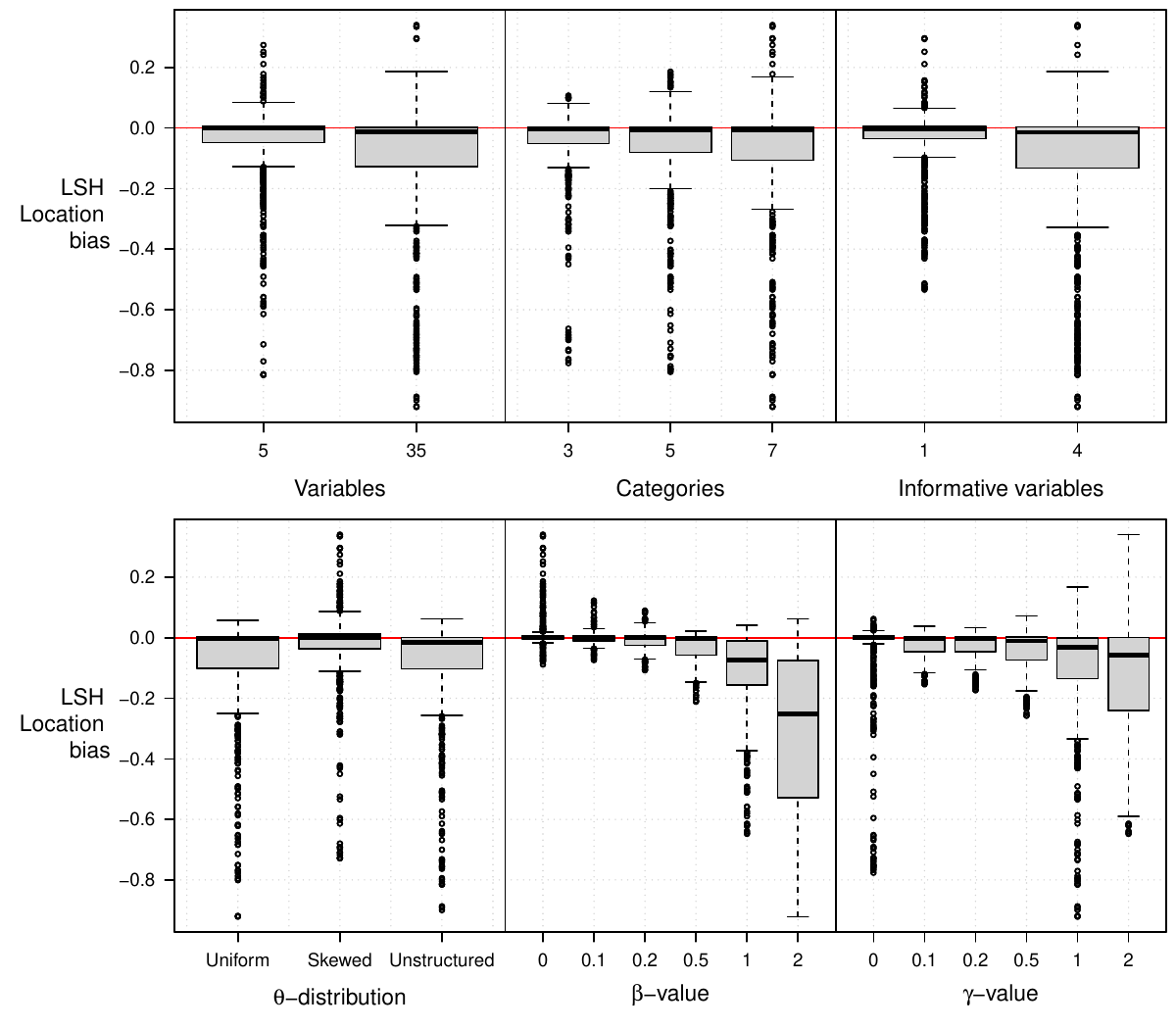}
    \caption{Influence of simulation settings on the bias of the location-shift model, when drawing with the location-shift model aggregated across all $n$. Except in their specific columns, boxplots aggregated across all numbers of covariates, number of categories, number of informative variables, true location and scale parameter values, and distribution settings.}
    \label{bias_loc_lsh}
\end{figure}

In Figure \ref{bias_loc_lsh}, biases for the location parameter of the location-shift model are displayed. The estimations of the location parameters showed increased dispersion with a larger number of covariates, compared to a lower number. Similarly, dispersion of the estimations increased with more categories and more informative covariates. A slightly larger bias could be seen for more informative covariates and the unstructured distribution. We saw that the higher the real value for the location parameter $\beta$, the higher the (negative) bias of the estimated location parameter, and high values were also accompanied by higher dispersion of the estimation. The (negative) bias of the location parameter also increased with increasing dispersion value, and high values were also accompanied by higher dispersion of the estimation.

\newpage

\begin{figure}[h]
    \centering
    \includegraphics[width=1\textwidth]{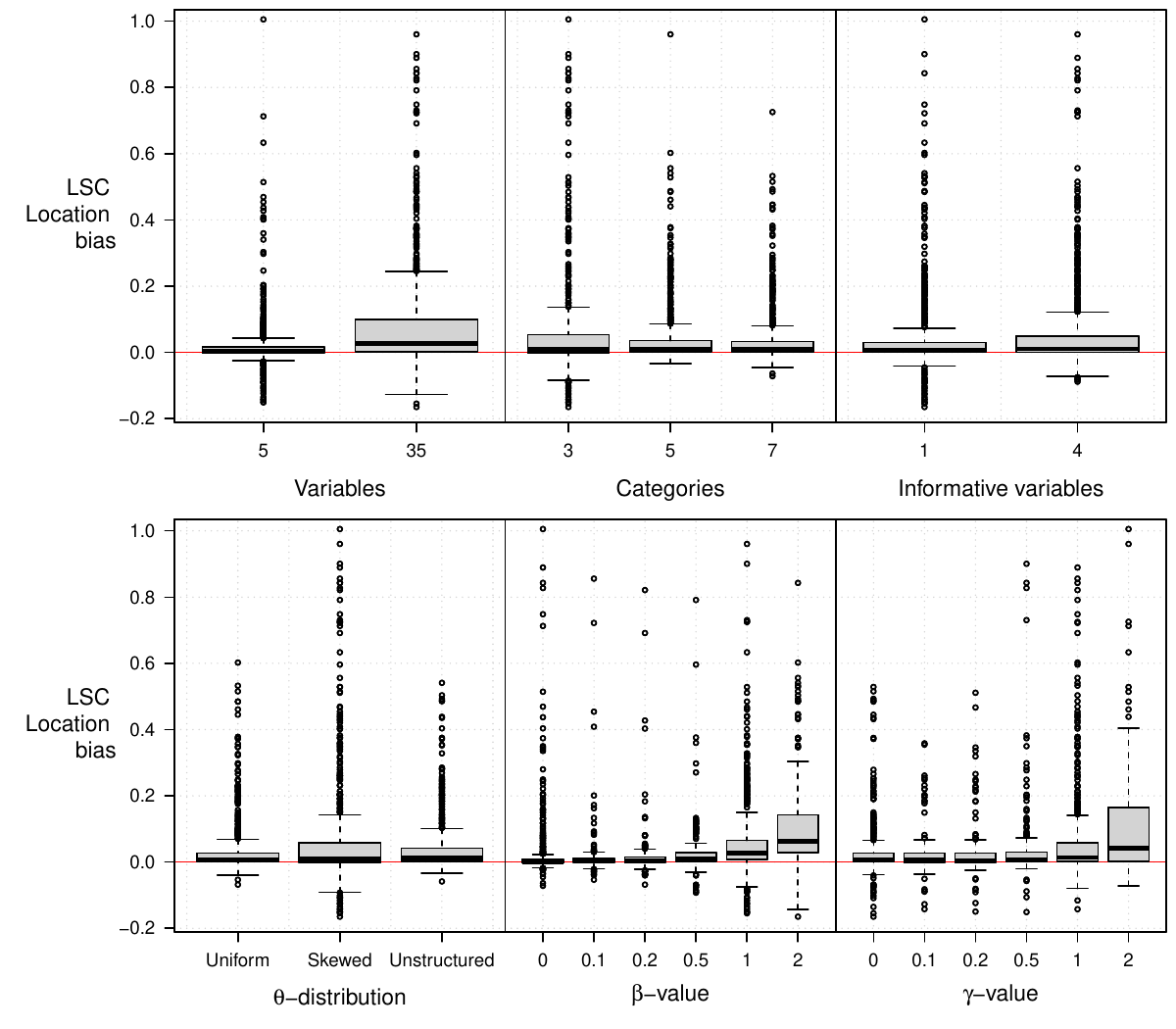}
    \caption{Influence of simulation settings on the bias of the location-scale model, when drawing with the location-scale model aggregated across all $n$. Except in their specific columns, boxplots aggregated across all numbers of covariates, number of categories, number of informative variables, true location and scale parameter values, and distribution settings. 12 values in the range of 1.04 to 2.49 were excluded from the Figures for the purpose of clearly seeing the boxes.}
    \label{bias_loc_lsc}
\end{figure}

In Figure \ref{bias_loc_lsc}, biases for the location parameter of the location-scale model are displayed. The estimations of the location parameters showed increased dispersion with a larger number of covariates, compared to a lower number. Similarly, dispersion of the estimations increased slightly with fewer categories and the skewed distribution. We saw that the higher the real value for the location parameter $\beta$, the higher the bias of the estimated location parameter, and high values were also accompanied by higher dispersion of the estimation. The bias of the location parameter also increased with increasing dispersion value, and high values were also accompanied by higher dispersion of the estimation.

\newpage

\newpage

\begin{figure}[h]
    \centering
    \includegraphics[width=1\textwidth]{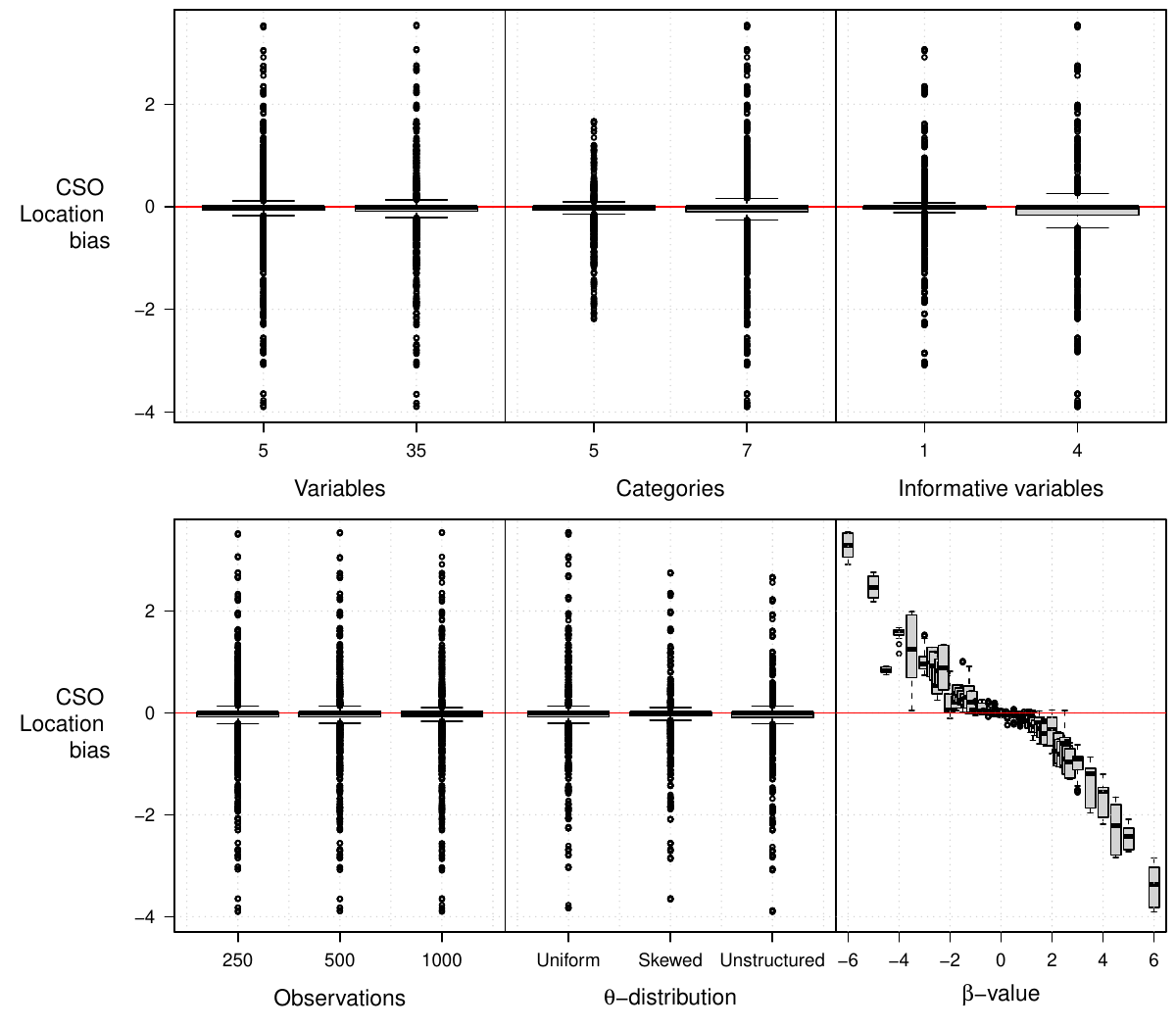}
    \caption{Influence of simulation settings on the bias of the category-specific odds model and models nested within the category-specific odds model. Except in their specific columns, boxplots aggregated across all numbers of observations, covariates, categories, number of informative variables, distributions, and true location parameter values.}
    \label{bias_cat_spec}
\end{figure}

In Figure \ref{bias_cat_spec}, biases for the location parameter of the category-specific odds model are displayed. As described in the Methods section, the location-shift model can be expressed as a category-specific odds model. Therefore, these models were included in the bias calculations as well, which lead to real $\beta$-values differing from the values chosen in the simulation settings, as the dispersion parameter needed to be expressed in the location parameters.

The estimations of the location parameters showed increased dispersion with a larger number of informative variables. Dispersion of the estimations seemed to increase with increasing categories, but this is due to larger $\beta$-values in settings with more categories when a bias term was drawn from the location-shift model. We saw that the higher the real value for the location parameter $\beta$, the higher the bias of the estimated location parameter, and high values were also accompanied by higher dispersion of the estimation. The number of variables, observations, and the distribution did not play a relevant role.

\section{Further results for the $\alpha$-errors}
\label{Further_res_alpha}

\begin{figure}[h]
    \centering
    \includegraphics[width=1\textwidth]{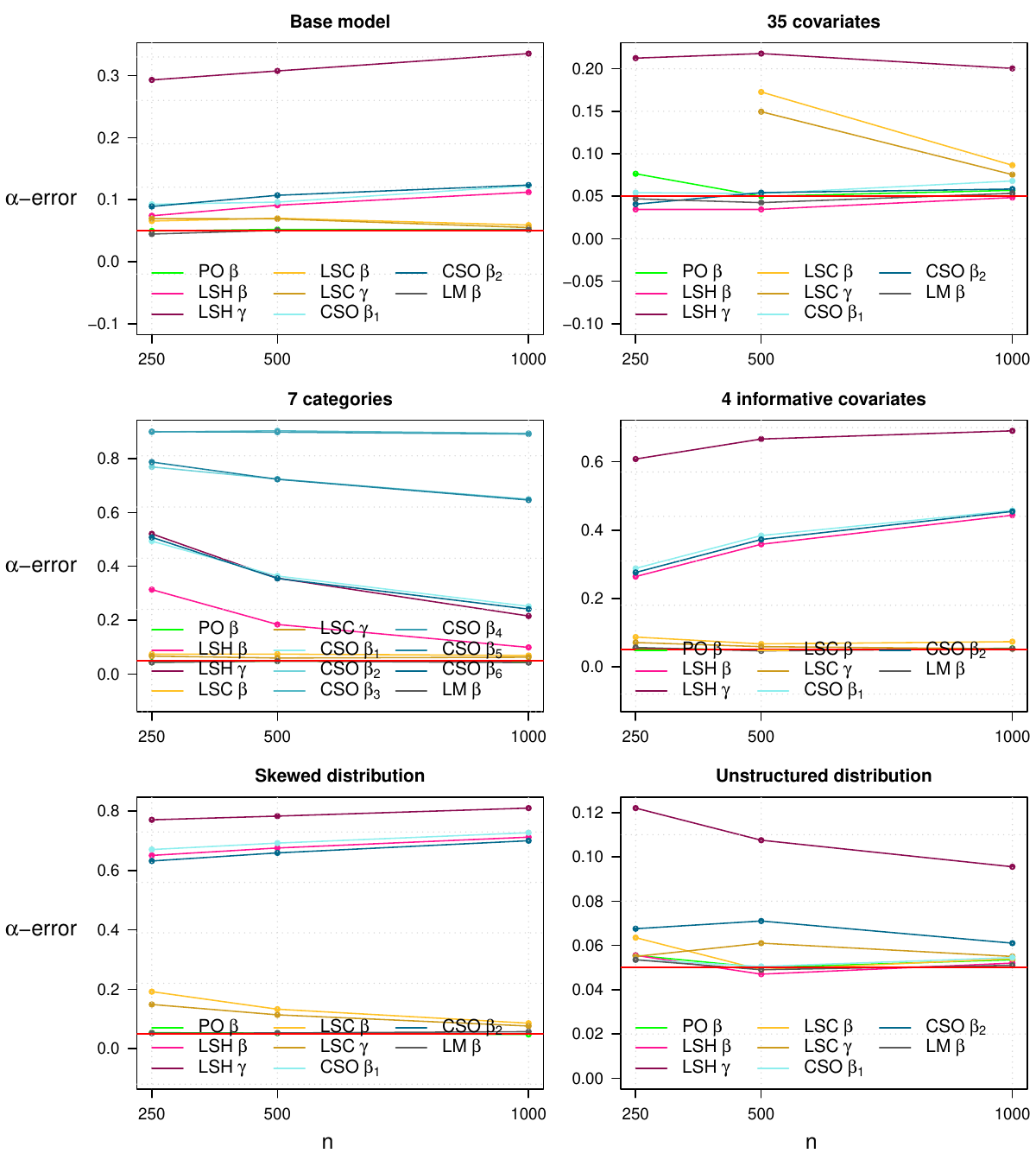}
    \caption{Influence of simulation settings on the $\alpha$-errors of the model. Data was drawn from the location-shift model with $\gamma = 1$ and $\beta = 0$}
    \label{alpha_disp_shift_v5_g1_b0}
\end{figure}

Figure \ref{alpha_disp_shift_v5_g1_b0} shows the $\alpha$-errors of an uninformative variable for all models when drawing with the location-shift model for $\gamma = 1$ and $\beta = 0$.

\newpage

\begin{figure}[h]
    \centering
    \includegraphics[width=1\textwidth]{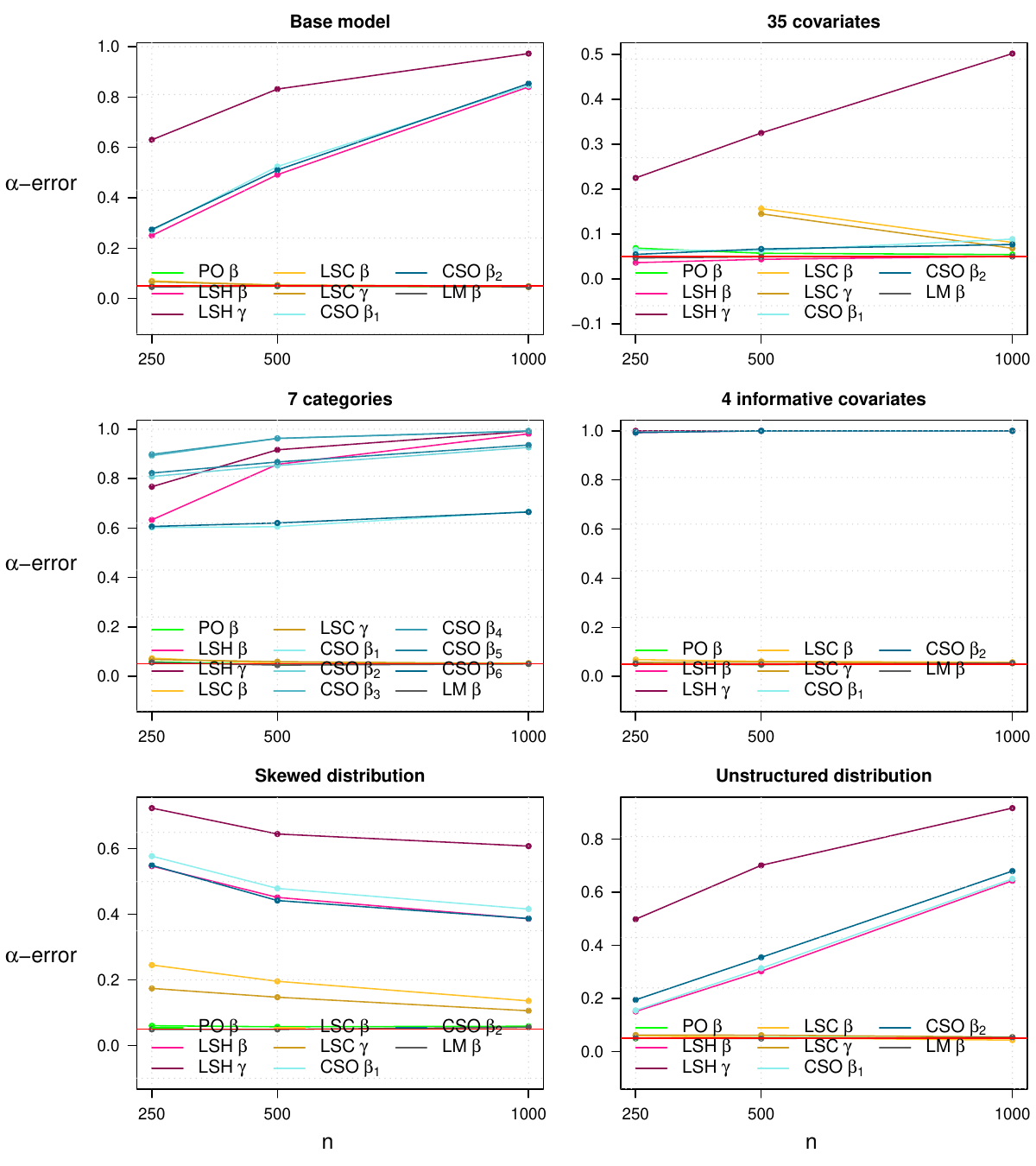}
    \caption{Influence of simulation settings on the $\alpha$-errors of the model. Data was drawn from the location-scale model with $\gamma = 1$ and $\beta = 0$}
    \label{alpha_disp_scale_v5_g1_b0}
\end{figure}

Figure \ref{alpha_disp_scale_v5_g1_b0} shows the $\alpha$-errors of an uninformative variable for all models when drawing from the location-scale model with $\gamma = 1$ and $\beta = 0$.

\newpage

\begin{figure}[h]
    \centering
    \includegraphics[width=0.95\textwidth]{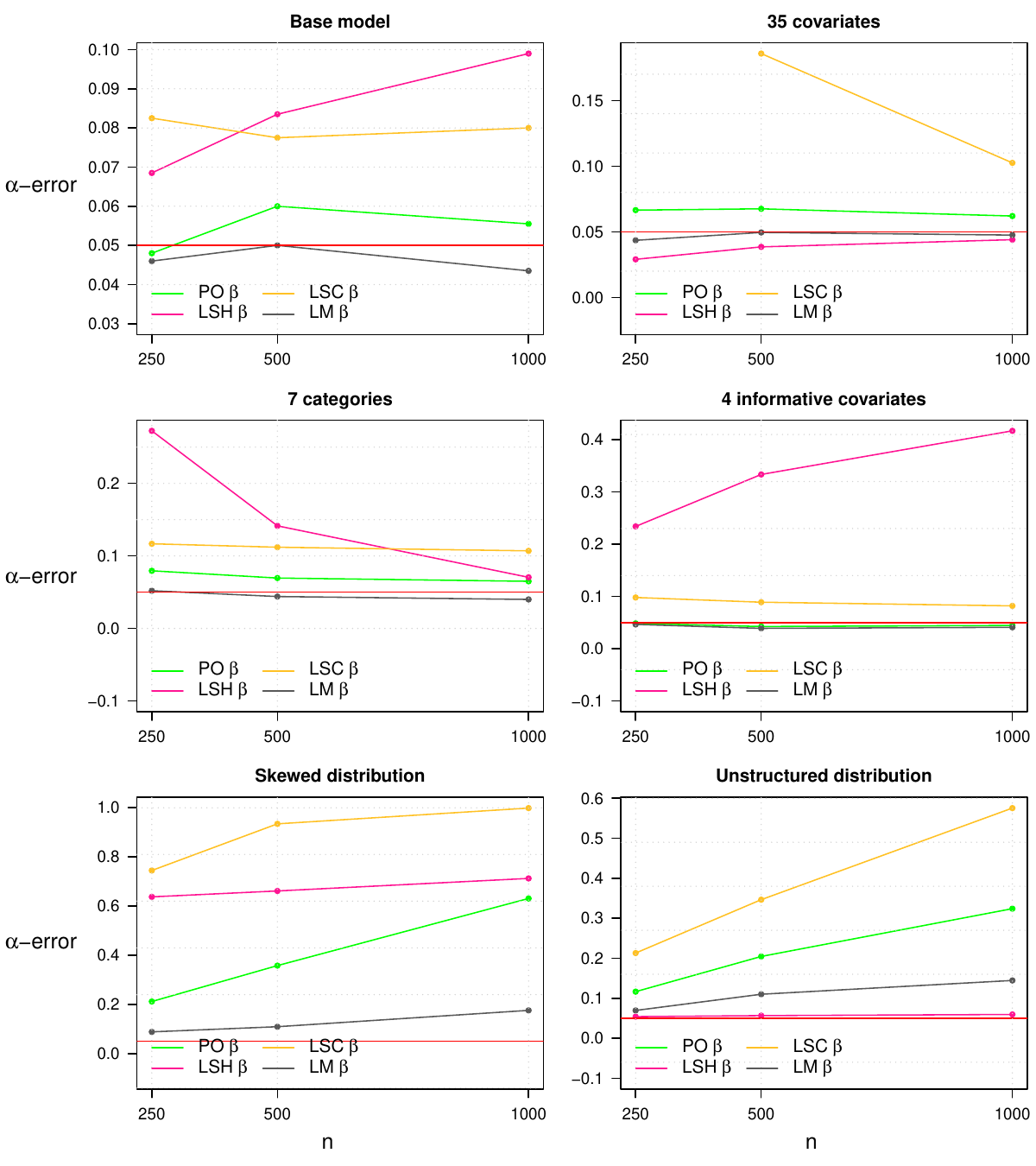}
    \caption{Influence of simulation settings on the $\alpha$-errors of the location parameter of the model. Data was drawn from the location-shift model with $\beta = 0$ and $\gamma = 1$. The chosen base model has $p = 5$ covariates, $k = 3$ categories, one informative covariate, and the uniform distribution setting.}
    \label{alpha_disp_shift_v1}
\end{figure}

Figure \ref{alpha_disp_shift_v1} shows the $\alpha$-errors of the location parameters for the linear model, the proportional odds model, and the two dispersion models of an informative variable when drawing from the location-shift model with $\gamma = 1$ and $\beta = 0$. The CSO model was not included as the location parameters are able to model dispersion, so the real $\beta$ values for the CSO are not zero.

\newpage

\begin{figure}[h]
    \centering
    \includegraphics[width=0.95\textwidth]{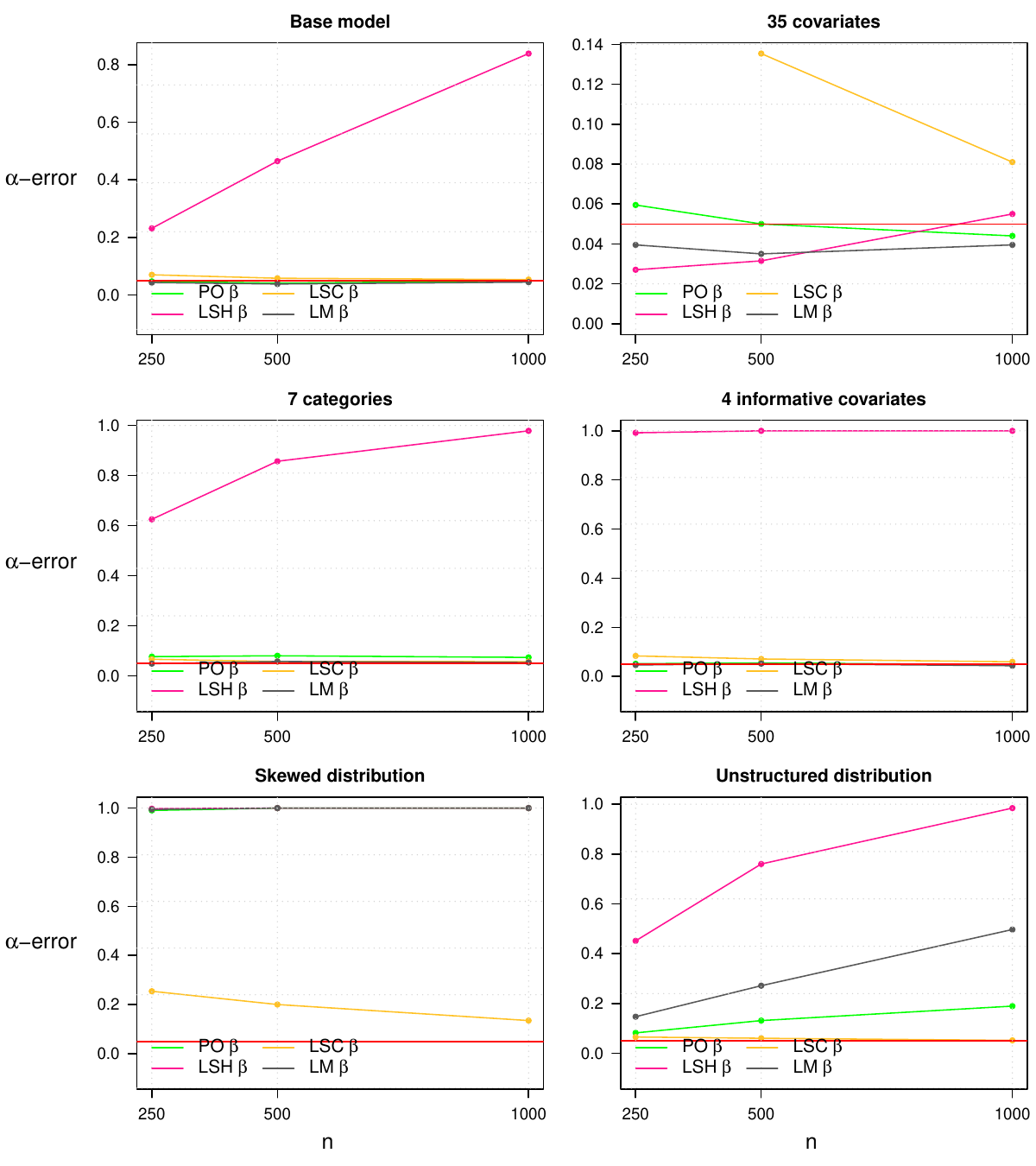}
    \caption{Influence of simulation settings on the $\alpha$-errors of the location parameter of the model. Data was drawn from the location-scale model with $\beta = 0$ and $\gamma = 1$. The chosen base model has $p = 5$ covariates, $k = 3$ categories, one informative covariate, and the uniform distribution setting.}
    \label{alpha_disp_scale_v1}
\end{figure}

Figure \ref{alpha_disp_scale_v1} shows the $\alpha$-errors of the location parameters for the linear model, the proportional odds model, and the two dispersion models of an informative variable when drawing from the location-scale model with $\gamma = 1$ and $\beta = 0$. The CSO model was not included as the location parameters are able to model dispersion, as in the case of the location-shift model, so the real $\beta$ values for the CSO may not be zero.

\newpage

\begin{figure}[h]
    \centering
    \includegraphics[width=0.95\textwidth]{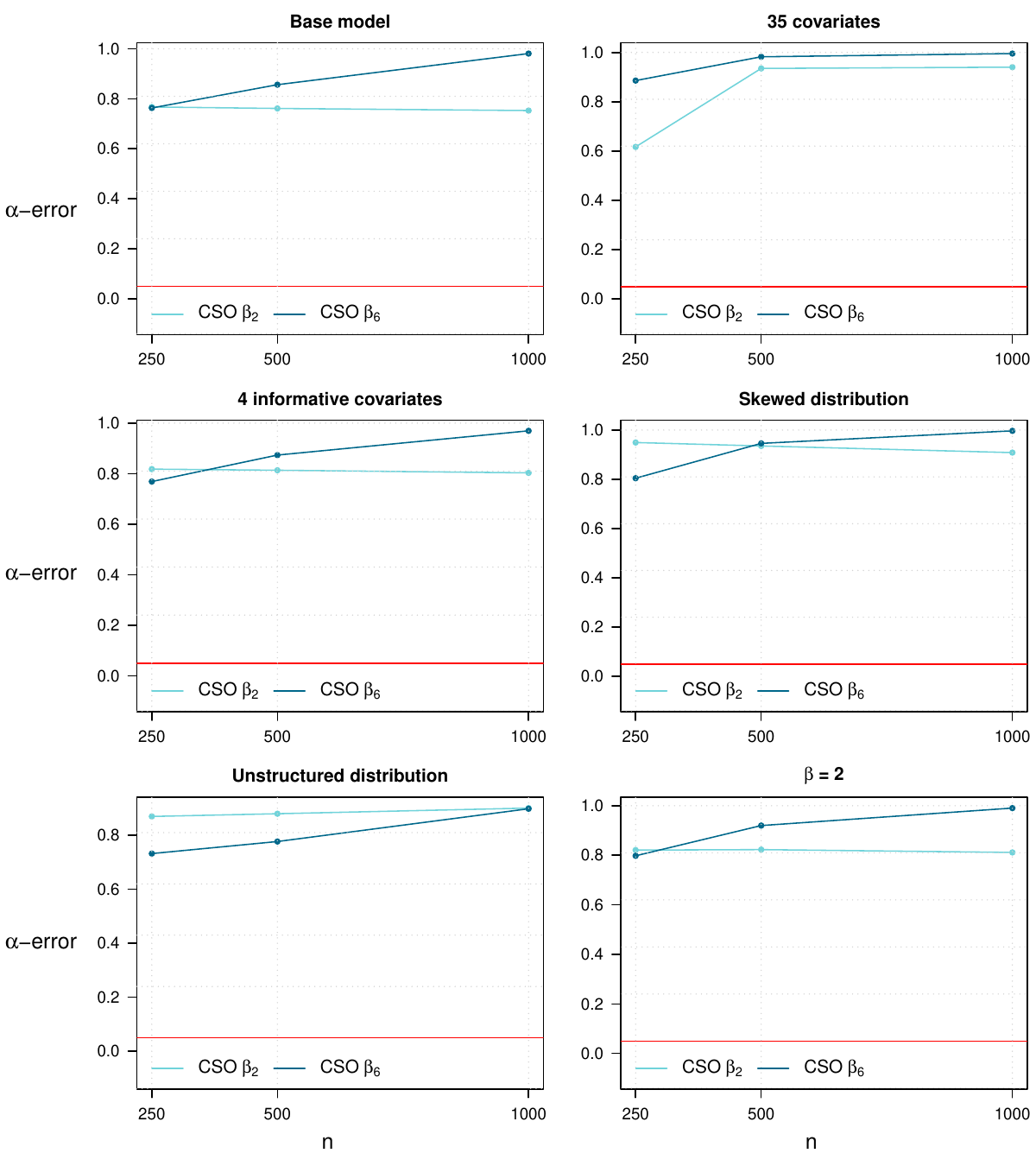}
    \caption{Influence of simulation settings on the $\alpha$-errors of the location parameter of the model. Data was drawn from the location-scale model with $\beta = 0$ and $\gamma = 1$. The chosen base model has $p = 5$ covariates, $k = 3$ categories, one informative covariate, and the uniform distribution setting.}
    \label{alpha_cso_vs_slight_vars}
\end{figure}

Figure \ref{alpha_cso_vs_slight_vars} displays results for the $\alpha$-error of the category-specific odds model parameters for data drawn from the CSO model. We present results for an informative covariate with $u = 1$, so that in the case of 7 covariates $\beta_2 = \beta_6 = 0$. 

\end{document}